\documentclass[fleqn,usenatbib]{mnras}

\usepackage{newtxtext,newtxmath}
\usepackage[T1]{fontenc}

\DeclareRobustCommand{\VAN}[3]{#2}
\let\VANthebibliography\thebibliography
\def\thebibliography{\DeclareRobustCommand{\VAN}[3]{##3}\VANthebibliography}

\usepackage{graphicx}	% Including figure files
\usepackage{amsmath}	% Advanced maths commands

\usepackage{amssymb}	% Extra maths symbols
\usepackage{bbold}      % used to display the unit tensor
\usepackage[normalem]{ulem}       %used for striked out text
\usepackage{commath}
\usepackage{bm}  
\usepackage{xcolor}

\title[3-D Dust Stirring by a Giant Planet]{Three-Dimensional Dust Stirring by a Giant Planet Embedded in a Protoplanetary Disk}

\author[F. Binkert et al.]{Fabian Binkert,$^{1,2}$\thanks{E-mail: fbinkert@usm.lmu.de}
 Judit Szul\'{a}gyi$^{3}$
and Til Birnstiel$^{1,2}$
\\
$^{1}$University Observatory, Faculty of Physics, Ludwig-Maximilians-Universität München, Scheinerstr. 1, 81679 Munich, Germany\\
$^{2}$Exzellenzcluster ORIGINS, Boltzmannstr. 2, D-85748 Garching, Germany\\
$^{3}$Institute for Particle Physics \& Astrophysics, ETH Zurich, Wolfgang-Pauli-Str. 27, 8093 Zürich, Switzerland}

\date{Accepted XXX. Received YYY; in original form ZZZ}

\pubyear{2022}

\begin{document}
\label{firstpage}
\pagerange{\pageref{firstpage}--\pageref{lastpage}}
\maketitle

\begin{abstract}
%Context
The motion of solid particles embedded in gaseous protoplanetary disks is influenced by turbulent fluctuations. Consequently, the dynamics of moderately to weakly coupled solids can be distinctly different from the dynamics of the gas. Additionally, gravitational perturbations from an embedded planet can further impact the dynamics of solids.
%Aims.
In this work, we investigate the combined effects of turbulent fluctuations and planetary dust stirring in a protoplanetary disk on three-dimensional dust morphology and on synthetic ALMA continuum observations. 
%Methods
We carry out 3D radiative two-fluid (gas+1-mm-dust) hydrodynamic simulations in which we explicitly model the gravitational perturbation of a Jupiter-mass planet. We derived a new momentum-conserving turbulent diffusion model that introduces a turbulent pressure to the pressureless dust fluid to capture the turbulent transport of dust. The model implicitly captures the effects of orbital oscillations and reproduces the theoretically predicted vertical settling-diffusion equilibrium. 
%Results
We find a Jupiter-mass planet to produce distinct and large-scale three-dimensional flow structures in the mm-size dust, which vary strongly in space. We quantify these effects by locally measuring an effective vertical diffusivity (equivalent alpha) and find azimuthally averaged values in a range $\delta_\mathrm{eff}\sim5\cdot 10^{-3} - 2\cdot 10^{-2}$ and local peaks at values of up to $\delta_\mathrm{eff}\sim3\cdot 10^{-1}$. In synthetic ALMA continuum observations of inclined disks, we find effects of turbulent diffusion to be observable, especially at disk edges, and effects of planetary dust stirring in edge-on observations. 
\end{abstract}


%%%%%%%%%%%%%%%%%%%%%%%%%%%%%%%%%%%%%%%%%%%%%%%%%%

%%%%%%%%%%%%%%%%% BODY OF PAPER %%%%%%%%%%%%%%%%%%
\section{Introduction}
Protoplanetary disks consist only of one per cent dust by mass. Even though the other 99 per cent is gaseous, it is the small solid component in protoplanetary disks from which all rocky objects, such as rocky planets, form. Advanced radio interferometers, such as the Atacama Large Millimeter/submillimeter Array (ALMA), are capable of detecting and resolving the faint thermal emission of cold dust in protoplanetary disks. This provides direct insight into the earliest phase of planet formation. Continuum observations of protoplanetary disks with ALMA have revealed numerous substructures, such as gaps, rings, and asymmetries \citep[see e.g. the review by][]{Andrews20}. Numerical studies predict a planet (or multiple planets) to be capable of producing many of the observed disk features via its gravitational interaction with the surrounding circumstellar disk \citep[e.g.][]{Wolf05,Gonzalez2012,Perez2015,Dong2015a,Dong2015b,Szulagyi2018,Szulagyi19,Weber20}. However, despite significant efforts, only a few observed disk substructures have been successfully linked to the presence of a planet \citep[e.g., PDS 70b/c,][]{Keppler2018,Isella19,Haffert19,Christiaens19}. \newline
Most theoretical and observational studies on disk substructures have focused on radial structures, favoring low-inclination disks because radial structure is more readily observable \citep[e.g.][]{ALMApartnership2015,Pinte15,Jin16,Dong2017a,Szulagyi2018,Dullemon18,Ricci18,Zhang2018,Andrews2018,Dipierro2018,Wafflard20}. Even though, it is difficult to constrain the vertical extent of the millimeter continuum emission in low-inclination disks due to their geometrically thin shape, the study of the vertical structure of more inclined and/or edge-on disks offers additional opportunities to constrain disk properties. \newline
%planetary stirring
The vertical extent of millimeter-sized dust grains, which are mainly probed with (sub-)millimeter continuum observations, is set by a balance of vertical settling and mixing. While small, micron-sized, dust grains are aerodynamically tightly coupled to their gaseous environment, the larger dust grains tend to decouple from the gas and settle towards the disk midplane, forming a geometrically thin midplane layer. However, these larger dust grains are still somewhat coupled to their environment and react to fluctuations in the gas. Thus, turbulent flows have the potential to counteract the vertical settling of moderately coupled dust grains, setting the vertical extent of the dust disk. Even though protoplanetary disks are generally found to be turbulent \citep[e.g.][]{Hughes11,Guilloteau12,Pinte15,Teague16,Dullemon18}, the driving mechanism of the turbulence is not yet fully understood despite the study of promising candidates as found in the magneto-rotational instability \citep[e.g.][]{Flock15} or in purely hydrodynamic mechanisms such as the vertical shear instability (VSI) \citep[e.g.][]{Urpin03,Nelson13,Stoll14,Schafer20}. In addition to disk turbulence, it has been shown that a planet, embedded in a protoplanetary disk, can be an additional source of dust mixing \citep[][]{Binkert21,Bi2021}. \newline
Due to the lack of full understanding regarding the origin of turbulence, the underlying driving mechanisms of turbulence are often disregarded when studying the dynamics of gas in protoplanetary disks using hydrodynamical simulations. Instead, the net effects of the unspecified turbulence on gas are parametrized with an effective turbulent viscosity \citep[][]{Shakura1973,Lynden-Bell1974}. On the other hand, the net effects of turbulent flows on dust have successfully been modeled by using a gradient-diffusion hypothesis, which models the turbulent mixing of dust grains in a turbulent environment \citep[][]{Cuzzi1993,Youdin2007,Carballido06,Carballido10,Zhu15}. The subsequent comparison between such hydrodynamical models and the observed radial structure and/or the vertical extent of circumstellar disk, allows for the constraint of the effective viscosity and/or diffusivity in the observed disks \citep[e.g.][]{Pinte15,Vilenave22}.\newline
Motivated by such comparisons, we build up on the results of our previous papers \citep[][]{Binkert21,Szulagyi2022}, in which we studied observational disk features caused by a planet, using radiative hydrodynamic two-fluid simulations (gas + millimeter-size dust). For the current work, we expand the dust module of the \textsc{Jupiter} code \citep[][]{Szulagyi2014} by treating turbulent diffusion as a pseudo pressure in the otherwise pressureless dust fluid, similar to the approach by \cite{Klahr2021}. Consequently, our approach ensures the conservation of angular and linear in our simulations, a property that traditional turbulent diffusion approaches lack \citep[][]{Tominaga19} and is crucial to correctly capture the dynamics of dust and gas within a circumstellar disk. In the following work, we show that our momentum-conserving diffusion approach implicitly captures the effects of orbital dynamics (both in-plane and vertical epicycles of dust grains \citep[][]{Youdin2007}, also when the dust flow deviates from a purely Keplerian flow, e.g., in the Hill sphere of an embedded planet. Further, we study the combined effects of turbulent mixing (parametrized by a diffusivity) and planetary mixing of millimeter-sized dust, on the global, three-dimensional dust distribution within a circumstellar disk. And, we study how turbulent stirring and planetary stirring affect the observed millimeter-continuum flux in synthetic ALMA observation of face-on and inclined disks. \newline
The outline of this paper is as follows. In section \ref{sec:method}, we describe the dynamical equations that govern dust, gas, and radiation in our hydrodynamic models. We also describe how we create synthetic ALMA continuum observations from these models. In section \ref{sec:turb_diff_in_dust}, we discuss the properties of our momentum-conserving turbulent diffusion model. In section \ref{sec:results}, we present our results and in section \ref{sec:Discussion}, we discuss and summarize our results. In the appendix, we present numerical benchmark tests of our diffusion model and compare our momentum-conserving diffusion model to the mass diffusing model.

%%%%%%%%%%%%%%%%%%%%%%%%%%%%%%%%%%%%%%%%%%%%%%%%%%%%%%%%%%%%%%%%%%%%%%%%%%%%%%%%%%%%%%%%%%%%%%%%%%%%
\section{Method}
\label{sec:method}
We run global three-dimensional radiative two-fluid (gas+mm-sized dust) hydrodynamic simulations of circumstellar disks with an embedded planet, to investigate the effect of planetary stirring on the three-dimensional dust morphology in the presence of turbulent diffusion. Further, we employ radiative transfer calculations to turn the hydrodynamic simulations into synthetic continuum intensity maps, from which we generate synthetic ALMA observations to study observational signatures of planetary stirring and/or background turbulent diffusion. This section is structured as follows. We introduce the physical models of dust and turbulent diffusion in section \ref{sec:diffusion_pressure_model}, and the hydrodynamic model of the gas and the radiation component in section \ref{sec:Gas and Radiation Model}. In section \ref{sec:method1}, we describe our simulation procedure and the numerical details. Sections \ref{sec:Radiative Transfer} and \ref{sec:syn_obs} describe the post-processing steps. Namely, the radiative transfer calculations and the subsequent generation of synthetic ALMA continuum observations, respectively. 
\subsection{Dust and Turbulent Diffusion Model}
\label{sec:diffusion_pressure_model}
Turbulent gas in protoplanetary disks contains a wide range of excited length scales, down to the molecular level. Dust grains embedded in these turbulent environments are aerodynamically coupled to the gas motion and, depending on their properties, can be excited on a similar range of lengths scales. Thus, capturing the entire dynamics of the dust-gas interactions in numerical hydrodynamic simulations would require spatial resolution down to the molecular dissipation scale of turbulence. However, this is, with today's computational resources, not possible in global hydrodynamical simulations of protoplanetary disks, and, the smallest dynamical scales remain unresolved. As a workaround, diffusion models have been adopted to model the effects of unresolved small-scale gas motions on resolved large-scale dust flows \citep[][]{Cuzzi1993,Fan1998}. Also in this work, we invoke the \textit{gradient diffusion hypothesis} and assume the resolved dust density flux to be driven by the gradient of the dust density $\rho_d$ itself and a transport coefficient $D$ (the so-called \textit{Fick's law}) as
\begin{equation}\label{eq:diffusion_flux_def}
    \mathbf{j} = -D\nabla \rho_d
\end{equation}
Above, the proportionality coefficient $D$ is called the \textit{dust diffusion coefficient}. Special care must be taken if dust turbulent diffusion happens in a non-uniform gas environment, for which equation (\ref{eq:diffusion_flux_def}) is usually modified. Here, we do not approach this complication before section \ref{sec:Stratified Gas Background} and assume the macroscopic quantities of the gas background which drives the turbulent diffusion to be uniform for now. The velocity associated with the dust diffusion flux $\mathbf{j}$ is the \textit{diffusion velocity}, which we define by dividing the diffusion flux by the dust density $\mathbf{v}_{\mathrm{diff}} = \mathbf{j} / \rho_d$. Thus, the explicit form of the diffusion velocity is
\begin{equation}\label{eq:def_diff_velocity}
    \mathbf{v}_{\mathrm{diff}}=-D\bm{\nabla}\ln\rho_d
\end{equation}
and describes a flow in the opposite direction of the gradient in the local dust density $\rho_d$. The value of the dust diffusion coefficient $D$ is determined by the turbulent properties of the gas. The gaseous component is generally modeled as a viscous fluid for which diffusive processes of unresolved turbulence are parametrized with an effective turbulent viscosity coefficient $\nu$. Both $D$ and $\nu$ have the same dimensionality and are related via the dimensionless Schmidt number $Sc$ \citep[][]{Cuzzi1993}:
\begin{equation}\label{eq:Sc_number}
    Sc=\frac{\nu}{D}
\end{equation}
{Throughout this work, we assume the dust diffusion coefficient $D$ to be constant in space}. In our model, we parametrize only the unresolved effects of subgrid turbulent motions with the diffusion coefficient. We assume these turbulent fluctuations to be isotropic, random, and correlated only on timescales shorter than any other relevant timescale in our models. \newline
To ensure mass conservation, the diffusion flux $\mathbf{j}$ is typically added to the continuity equation as a second contribution to advection besides the regular advection flux $\rho_d\mathbf{v}_d$, where $\mathbf{v}_d$ is the resolved dust advection velocity. The dust continuity equation then has the following form:
\begin{equation}\label{eq:new_dust_adv_equation}
\frac{\partial \rho_d}{\partial t}+\bm{\nabla}\cdot\Big(\rho_d( \mathbf{v}_d+\mathbf{v}_{\mathrm{diff}})\Big)=0
\end{equation}
Motivated by equation (\ref{eq:new_dust_adv_equation}), we define an \textit{effective dust advection velocity} $\mathbf{v}^{*}_{d}$ as a sum of the two velocity components
\begin{equation}\label{eq:comb_velocity}
      \mathbf{v}^{*}_{d} =  \mathbf{v}_{d}+\mathbf{v}_{\mathrm{diff}}
\end{equation}
The first component is the resolved dust advection velocity $\mathbf{v}_{d}$, which describes the contribution from resolved processes to the effective advection velocity of the dust $\mathbf{v}^{*}_{d}$. The second component  $\mathbf{v}_{\mathrm{diff}}$ is the resolved mean contribution from subgrid interactions of the dust with the turbulent gas to the effective advection velocity $\mathbf{v}^{*}_{d}$. In the absence of subgrid turbulence, the second term in equation (\ref{eq:comb_velocity}) is trivially zero and the effective dust advection velocity is equal to the resolved advection component $\mathbf{v}^{*}_{d} =  \mathbf{v}_{d}$. The continuity equation in terms of the effective dust advection velocity simplifies to
\begin{equation} \label{eq:dust_cont_eq2}
\frac{\partial \rho_d}{\partial t}+\bm{\nabla}\cdot(\rho_d \mathbf{v}^{*}_d)=0
\end{equation}
\newline
Next, we are interested in the dynamics of the effective dust advection velocity and consider the material derivative of the effective dust advection velocity and write it in Eulerian form with appropriate source terms as follows:
\begin{equation}\label{eq:vel_dash_definition}
\frac{\partial \mathbf{v}^{*}_d}{\partial t}+\big(\mathbf{v}^{*}_d\cdot\bm{\nabla}\big)\mathbf{v}^{*}_d=-\nabla\Phi-\frac{1}{\tau_s}(\mathbf{v}_d-\mathbf{v}_g)
\end{equation}
The first term on the r.h.s. of equation (\ref{eq:vel_dash_definition}) is the gravitational acceleration which is a result of an external force and thus acts on all spatial scales, i.e., both the resolved and diffusion flow components $\mathbf{v}_d$ and $\mathbf{v}_\mathrm{diff}$. In this term, $\Phi$ is the gravitational potential. The second term on the r.h.s. of equation (\ref{eq:vel_dash_definition}) is the explicit contribution from aerodynamic drag which acts on the resolved component of the flow, i.e., the velocity component $\mathbf{v}_d$, and is parametrized with the stopping time $\tau_s$ \citep{Whipple1972,Weidenschilling1977}. We assume Epstein drag, which is valid as long as the dust particles are smaller than the mean free path of the gas ($a<\lambda_\mathrm{mfp}$), and is valid for the range of disk parameters and the dust particle size assumed in this study. We write the stopping time of a dust grain of radius $a$ explicitly as 
\begin{equation}
\label{eq:stopping_time}
\tau_s=\sqrt{\frac{\pi}{8}}\cdot\frac{\rho_{\bullet} a}{\rho_g c_s}
\end{equation} 
The stopping time in equation (\ref{eq:stopping_time}) is larger, i.e., the dust-gas coupling is weaker, for dust particles with a large solid density $\rho_{\bullet}$, which depends on the composition of the dust grains. On the other hand, the stopping time is smaller, i.e., the coupling is stronger, if the dust grains are suspended in gas which has a larger (isothermal) sound speed $c_s=\sqrt{k_B T_g/m_{\mu}}$ with $T_g$ being the gas temperature and $m_\mu=3.9\times10^{-24}$ g the mean mass of a gas molecule. In a protoplanetary disk environment, we find the dimensionless \textit{Stokes number} by multiplying the stopping time with the local Keplerian frequency $\Omega$
\begin{equation}
    St=\tau_s\Omega
\end{equation}
Even though aerodynamic forces act on all spatial scales between the molecular scale and the system scale (i.e., resolved and subgrid scales), the drag term in equation (\ref{eq:vel_dash_definition}) acts only on the resolved velocity component $\mathbf{v}_d$ and not on $\mathbf{v}^{*}_d$ because the net effect of subgrid aerodynamic interactions between the dust and gas are already phenomenologically modeled with the definition of the  diffusion velocity in equation (\ref{eq:def_diff_velocity}). Hence, an explicit drag term acting on the diffusion velocity $\mathbf{v}_\mathrm{diff}$ does not need to be added to equation (\ref{eq:vel_dash_definition}).\newline
We now fully describe the dynamics of the dust fluid with the variables $\rho_d$ and $\mathbf{v}^{*}_d$, and equations (\ref{eq:dust_cont_eq2})  and (\ref{eq:vel_dash_definition}). However, the fact that two different dust velocity components appear in equation (\ref{eq:vel_dash_definition}) does not make the expression very intuitive and inconvenient to solve numerically. Thus, we rewrite equation (\ref{eq:vel_dash_definition}) in terms of the effective dust advection velocity $\mathbf{v}^{*}_d$ only, using definitions in equations (\ref{eq:def_diff_velocity}) and (\ref{eq:comb_velocity}):
\begin{equation} \label{eq:velocity_equation_with_diff03}
\frac{\partial \mathbf{v}^{*}_d}{\partial t}+\big(\mathbf{v}^{*}_d\cdot\bm{\nabla}\big)\mathbf{v}^{*}_d= -\nabla\Phi-\frac{1}{\tau_s}(\mathbf{v}^{*}_d-\mathbf{v}_g)-\frac{D}{\tau_s}\frac{\nabla \rho_d}{\rho_d}
\end{equation}
We identify equation (\ref{eq:velocity_equation_with_diff03}) to be formally equivalent to the velocity equation of a gas fluid with an isothermal equation of state with sound speed $\sqrt{D/\tau_s}$. Interestingly, equation (\ref{eq:velocity_equation_with_diff03}) is identical to the expression introduced in appendix B of \cite{Klahr2021} who use, in their derivation, a settling-diffusion equilibrium ansatz similar to the derivation of Brownian motion used by \cite{Einstein1905}, to arrive at the expression. In their work, \cite{Klahr2021} call the third term on the r.h.s. in equation (\ref{eq:velocity_equation_with_diff03}) \textit{diffusive pressure} due to its functional similarity to the gas pressure. We will also adopt this term here. The diffusive pressure term acts to smear out gradients in the dust density, similar to the gas pressure, which acts to expand the gas. It is now also more intuitive to understand that turbulent diffusion drives the evolution of the dust fluid via a pressure-like term in the velocity equation, while the drag term acts on all the velocity components. 
\newline
We define the \textit{diffusion pressure}, i.e., the pseudo pressure that arises in the pressureless dust fluid as a result of turbulent diffusion, as 
\begin{equation}\label{eq:diff_press_def}
    P_d = \rho_dc_d^2
\end{equation}
with the \textit{diffusion speed} $c_d$ defined as 
\begin{equation}\label{eq:def_pebble_speed}
    c_d^2 = \frac{D}{\tau_s}
\end{equation}
(note, \cite{Klahr2021} call this the \textit{pebble speed}) Thus, we write the final system of coupled equations that describe the dynamics of the dust fluid in conservation form, using the definition of the diffusion speed $c_d$, as 
\begin{equation} \tag{\ref{eq:dust_cont_eq2}}
\frac{\partial \rho_d}{\partial t}+\bm{\nabla}\cdot(\rho_d \mathbf{v}^{*}_d)=0
\end{equation}
\begin{equation}
\label{eq:momentum_conservation_03}
\begin{split}
    \frac{\partial}{\partial t} ( \rho_d \mathbf{v}^{*}_d)+\bm{\nabla}\cdot\bigg(\rho_d\: \mathbf{v}^{*}_d \otimes\mathbf{v}^{*}_d +\frac{1}{3} \rho_d c_d^2  \mathbb{1} \bigg)= 
    &-\rho_d\nabla\Phi \\
    &-\frac{\rho_d}{\tau_s}(\mathbf{v}^{*}_d-\mathbf{v}_g)
    \end{split}
\end{equation}
where equation (\ref{eq:dust_cont_eq2}) is the continuity equation, and equation (\ref{eq:momentum_conservation_03}) is the momentum equation in conservation form. Now it is apparent that, by describing the dust dynamics with the effective advection velocity $\mathbf{v}^{*}_d$, the momentum equation (\ref{eq:momentum_conservation_03}) does not represent a pressureless fluid anymore, but it represents a momentum equation of a fluid with pressure $\rho_dc_d^2$. The terms on the r.h.s. of equation (\ref{eq:momentum_conservation_03}) account for gravity and aerodynamic drag, respectively. 
{Looking at the conservation form of the momentum equation (\ref{eq:momentum_conservation_03}), it is apparent that, ignoring gravitational contributions, our approach conserves linear momentum in the dust fluid unless momentum is exchanged with the gas via the drag term. Following the steps of \cite{Tominaga19}, equation (\ref{eq:momentum_conservation_03}) can be rewritten to also make the conservation of angular momentum ($r\sin \theta \rho_dv^*_{d,\phi}$) apparent.} {In section \ref{sec:Stratified Gas Background} and section \ref{sec:Gas and Radiation Model}, respectively, we will describe that, in a stratified gas background, turbulent diffusion exchanges momentum between the dust and gas fluids and momentum is only conserved in the full system (dust + gas).} \newline

\subsubsection{Turbulent Diffusion in a Stratified Gas Background}
\label{sec:Stratified Gas Background}
So far, we have only considered dust turbulent diffusion in a uniform gas background (i.e., $\tau_s$ = const.), for which the turbulent dust mass flux as defined in equation (\ref{eq:diffusion_flux_def}) is applicable without restrictions. Now, we consider a nonuniform background in which the stopping time $\tau_s$ varies in space. This ultimately also applies to protoplanetary disks. Following our derivation of the previous sections, we find that spatial variations in the stopping time introduce an additional third term on the r.h.s of the dust momentum equation (\ref{eq:momentum_conservation_03})
\begin{equation} \label{eq:momentum_equation_4}
    \frac{\partial}{\partial t} ( \rho_d \mathbf{v}^{*}_d)+...=...\mathbin{+\rho_d\nabla c_d^2 }
\end{equation}
$\tau_s$ {This new term favors dust momentum transport towards the gradient of the diffusion speed $c_d^2$, or equivalently, according to the definition in equation (\ref{eq:def_pebble_speed}), toward decreasing values of the stopping time in stratified gas backgrounds}. \newline
If the nonuniform background arises from a gradient in the gas density, we must also account for a turbulent diffusion flux in the gas, which introduces a systematic diffusion velocity to the gas. For well-coupled dust-gas mixtures, the gas diffusion velocity must be identical to the dust diffusion velocity, therefore
\begin{equation}
    \mathbf{v}_\mathrm{diff,g}=-D\nabla \ln \rho_g
\end{equation}
where the diffusion coefficient $D$ in the dust and gas are identical. So far, in our formalism, this systematic velocity is contained in the velocity $\mathbf{v}_g$. However, following a similar argument made previously for the dust, the explicit drag term in our equations does not act on the diffusion component of the flow. Therefore, in a nonuniform gas background, we must consider this contribution in our momentum equation and the equation then reads 
\begin{equation}\label{eq:momentum_conservation_04}
\begin{split}
    \frac{\partial}{\partial t} (\rho_d \mathbf{v}^{*}_d)+\bm{\nabla}\cdot\bigg(& \rho_d\: \mathbf{v}^{*}_d \otimes\mathbf{v}^{*}_d +\frac{1}{3} \rho_d c_d^2  \mathbb{1} \bigg)=-\rho_d\nabla\Phi \\
    &-\frac{\rho_d}{\tau_s}(\mathbf{v}^{*}_d-\mathbf{v}_g)+\frac{\rho_d}{\rho_g}\nabla\Big(\rho_g c_d^2\Big)
    \end{split}
\end{equation}
It is interesting to note that, we would arrive at the same momentum equation (\ref{eq:momentum_conservation_04}) by initially adding a term of the form $D\nabla \ln \rho_g$ to the diffusion flux in equation (\ref{eq:diffusion_flux_def}). This approach has been taken frequently in the literature \citep[e.g.][]{Dubrulle1995,Schrapler04,DullemondD04}, but must be motivated differently. \newline

\subsection{Gas and Radiation Model}
\label{sec:Gas and Radiation Model}
The radiative gas model in this study is identical to the one used in \cite{Szulagyi2016} and \cite{Binkert21}. Particularly, we model the gas with an adiabatic equation of state and a radiative transfer module to account for heating (adiabatic heating, viscous heating, stellar irradiation) and cooling (adiabatic cooling, radiative cooling). The mass- and momenta equations in conservation form are:
\begin{equation}\label{eq:gas_continuity}
\frac{\partial \rho_g}{\partial t}+\bm{\nabla}\cdot(\rho_g \mathbf{v}_g)=0
\end{equation}
\begin{equation}
\label{eq:gas_momentum_conservation}
\begin{split}
    \frac{\partial}{\partial t} ( \rho_g \mathbf{v}_g)+\bm{\nabla}\cdot\bigg(& \rho_g\: \mathbf{v}_g \otimes\mathbf{v}_g +\frac{1}{3} P_g \mathbb{1} \bigg)= 
    -\rho_g\nabla\Phi\\
    &+\bm{\nabla}\cdot\bar{\bar{\tau}}
    -\frac{\rho_d}{\tau_s}(\mathbf{v}_g-\mathbf{v}^{*}_d)-\frac{\rho_d}{\rho_g}\nabla\Big(\rho_g c_d^2\Big)
    \end{split}
\end{equation}
Here, $\rho_g$ is the volume density of the gas, and $\mathbf{v}_g$ is its velocity vector. The gas pressure $P_g$ is coupled to the internal energy of the gas $\epsilon$ via the adiabatic equation of state:
\begin{equation}\label{eq:of_state}
    P_g=(\gamma-1)\epsilon
\end{equation}
where $\gamma=1.43$ is the adiabatic index. The first term on the r.h.s. of the momentum equation (\ref{eq:gas_momentum_conservation}) accounts for the change in momentum due to gravitational acceleration. The {third} term accounts for the momentum exchange with the dust, i.e., the back reaction, due to aerodynamic drag. Compared to the equation without dust turbulent diffusion in \cite{Binkert21}, it contains an additional term on the r.h.s. to account for the drag interaction on the diffusion flux. The source term exactly cancels with the corresponding source term in the dust momentum equation, which ensures the conservation of momentum in the full system (gas+dust). It also becomes apparent that diffusion, like the explicit drag, in this formulation, has a back reaction from the dust onto the gas. The second term on the r.h.s. of the momentum equation contains the stress tensor $\bar{\bar{\tau}}$, which is defined as
\begin{equation}
    \bar{\bar{\tau}}=2\rho_g\nu\bigg(\bar{\bar{\Gamma}}-\frac{1}{3}\big(\nabla\cdot \mathbf{v}_g) \mathbb{1}\big)\bigg)
\end{equation}
where $\bar{\bar{\Gamma}}$ is the strain tensor and $\nu$ is the kinematic viscosity. The third conservation equation that we solve is the energy equation that governs the evolution of the total energy (internal and kinetic energy). {We assume the thermal internal energy of the dust fluid to be zero at all times, thus, the energy equation describes the total energy of the gas ($E$) only}: 
\begin{equation}
\label{eq:energy_conservation}
\begin{split}
    \frac{\partial E}{\partial t}+\bm{\nabla}\cdot\bigg(\big( P_g \mathbb{1}-\bar{\bar{\tau}} \big) \cdot \mathbf{v}_g+E\mathbf{v}_g\bigg)=
    &-\rho_g\mathbf{v}_g\cdot\nabla\Phi + S\\
    &- \rho_g\kappa_P\big(B(T)-c\epsilon_\mathrm{rad}\big) 
    \end{split}
\end{equation}
Besides the term accounting for advection, the terms on the l.h.s. of equation (\ref{eq:energy_conservation}) containing the pressure $P_g$ and the stress tensor $\bar{\bar{\tau}}$ that accounts for adiabatic heating/cooling and viscous heating respectively. The first term on the r.h.s. of equation (\ref{eq:energy_conservation}) accounts for the work done by gravity, and the second term is the contribution from stellar heating. The last term on the r.h.s. of equation (\ref{eq:energy_conservation}) accounts for radiative heating/cooling, where $B(T)$ describes the total emitted power of a blackbody at temperature $T$, $\kappa_P$ is the Planck  opacity, and $c$ the speed of light. There is radiative cooling if the gas radiates more energy than it receives from the surrounding radiation field ($B(T)>c\epsilon_\mathrm{rad}$). The gas is radiatively heated if it receives more energy from the radiation field than it emits ($B(T)<c\epsilon_\mathrm{rad}$). The gas is in local thermodynamic equilibrium if the two terms balance each other, i.e.,  $B(T)=c\epsilon_\mathrm{rad}$ (assuming no other heating/cooling mechanisms are active). A fourth partial differential equation (PDE) describes the rate of change of the radiative energy density $\epsilon_\mathrm{rad}$:
\begin{equation}
\label{eq:radiation_energy}
    \frac{\partial \epsilon_\mathrm{rad}}{\partial t}=-\nabla\cdot \mathbf{F}_\mathrm{rad}+\rho_g\kappa_P \big( B(T)-c\epsilon_\mathrm{rad} \big)
\end{equation}
where the second term on the r.h.s. of equation (\ref{eq:radiation_energy}) is identical to the third term of equation (\ref{eq:energy_conservation}) and accounts for the contribution to the radiative energy from thermal emission and/or absorption of the gas. The first term on the r.h.s. of equation (\ref{eq:radiation_energy}) contains the radiative flux $\mathbf{F}_\mathrm{rad}$, which we find using the flux-limited diffusion approximation  \citep[see e.g.][]{Szulagyi2016}. {Specifically, the radiative flux can be expressed as}
\begin{equation}
    \mathbf{F}_\mathrm{rad} = -\frac{c \lambda}{\rho \kappa_R}\nabla \epsilon_\mathrm{rad}
\end{equation}
{where $\lambda$ is the flux limiter of \citet{Kley89}, and $\kappa_R$ is the Rosseland mean opacity.} The latter is a weighted average over frequency, defined as 
\begin{equation}
    \kappa_R(T,\rho_g)^{-1} = \frac{\int \kappa_{\nu}^{-1}(T,\rho_g)\frac{\partial B_\nu(T)}{\partial T}\mathrm{d}\nu}{\int \frac{\partial B_\nu(T)}{\partial T}\mathrm{d}\nu}
\end{equation}
{where $\kappa_{\nu}$ is the frequency dependent absorption opacity. More details about the opacities used here are given in section \ref{sec:method1}.}\newline
Ultimately, we calculate the gas temperature $T_g$ self-consistently with
\begin{equation}
    T_g = (\gamma-1)\frac{m_{\mu}\epsilon}{k_B\rho_g}.
\end{equation}
{We do not calculate the dust temperature during the hydrodynamic simulations because the thermal internal energy of the dust is assumed zero and consequently the dust temperature does not impact the dynamics of the dust fluid. The presence of the dust only implicitly affects the gas temperature via the opacity.} Ultimately, the total system of coupled equations to solve for the gas and radiation components are equations (\ref{eq:gas_continuity}), (\ref{eq:gas_momentum_conservation}), (\ref{eq:energy_conservation}) and (\ref{eq:radiation_energy}) which are in turn coupled to the dust equations (\ref{eq:dust_cont_eq2}) and (\ref{eq:momentum_conservation_03}) via the aerodynamic drag and turbulent diffusion terms. {It is important to highlight that, in contrast to the dust fluid, we have not introduced a turbulent pressure term to the gas momentum equation (\ref{eq:gas_momentum_conservation}), nor to the energy equation (\ref{eq:energy_conservation}) under the assumption that the term is small compared to the thermal pressure $P_g$.}

\subsection{Hydrodynamic Simulations}
\label{sec:method1}
As this work is a continuation of \cite{Binkert21}, we base the set of hydrodynamic simulations carried out in this paper, on the set used in \cite{Binkert21} and use identical setups and parameters but with the addition of turbulent diffusion. In particular, we run three-dimensional radiative two-fluid (gas+dust) hydrodynamic simulations of circumstellar disks with an embedded planet. The simulations are carried out with the grid-based code \textsc{Jupiter} \citep{Szulagyi2016} which solves the radiation and gas hydrodynamic equations, summarized in section \ref{sec:Gas and Radiation Model}, and are fully described in \cite{Szulagyi2016}. We modified the dust-solver of the \textsc{Jupiter} code, introduced in \cite{Binkert21}, to include the effects of turbulent diffusion on the dust fluid as described in section \ref{sec:diffusion_pressure_model}. As a result, we model the effects of subgrid turbulence on the dust via a dynamical diffusion pressure that ensures the conservation of angular and linear momentum in the system. \newline
In our simulations, we fix the dust grain size at $a=1$ mm throughout this entire work. As a result, the Stokes number, i.e., the degree of dust-gas coupling, freely changes depending on the local hydrodynamic conditions. \newline
{Further, we set the Planck opacity equal to the Rosseland mean opacity $\kappa_P= \kappa_R$ in favor of a shorter computational time. The difference between the two opacities is not large, and thus our results are not affected by this approximation \citep{Semenov03,Bitsch2013}. Henceforth, we drop the subscript $R$ and only consider the frequency averaged gas opacity $\kappa(T,\rho_g)$ that is a function of the local gas density and temperature.\newline
At temperatures below 1500~K, we assume dust to be the dominant contributor to the opacity, which is a valid approximation in the wavelength regime relevant for hydrodynamic heating and cooling. Further, we assume the gas to be thermally coupled to the dust. Based on these two assumptions, we calculate the frequency-dependent opacity $\kappa_\nu$ of three dust compounds (silicate, water ice, carbonaceous material) self-consistently with a version of the \textsc{bhmie} code of \cite{Bohren1984}. We then calculate a mass-weighted average of the individual Rosseland mean opacities for a combined dust composition of 40 per cent silicates, 40 per cent water ice, and 20 per cent carbonaceous material \citep[][]{Zubko1996, Draine03, Warren08}, and a dust-to-gas ratio of $\rho_d/\rho_g=0.01$. \newline
Above, 1500~K, the opacity $\kappa$ includes gas opacities from \citet{BellLin94}.} In detail, the implemented opacity table accounts for the sublimation of water ice (170 K), carbonaceous material (1500 K), and silicate (2000 K) respectively. Above 2000 K, only gas opacities contribute. {We refer to \citet{Szulagyi19} for more details on the construction of the opacity table. \newline
In our two-fluid setup, we modify the frequency averaged opacity $\kappa(T,\rho_g)$ compared to the one-fluid setup in \citet{Szulagyi2018}, such that for $T<1500$~K, it also includes the local dust density $\rho_d$. The two-fluid opacity $\kappa_\textrm{2f}$ that we ultimately implemented is calculated as:} 
\begin{equation}\label{eq:twi_fluid_opacity}
    \kappa_\textrm{2f}(T,\rho_g,\rho_d) = \kappa (T,0.01\rho_g+99\rho_d)
\end{equation}
{Note that for a local dust-to-gas ratio of 0.01, the two opacities are identical $\kappa_\textrm{2f}=\kappa$.} \newline
As a result of the above definition (equation \ref{eq:twi_fluid_opacity}), the dust-to-gas ratio is not fixed at one per cent, and regions with a large dust-to-gas ratio are more optically thick in our radiative simulations, while regions with a small dust-to-gas ratio are more optically thin. \newline
{In computational cells which are directly irradiated by stellar irradiation, i.e., cells in the disk surface, the opacity should not depend on the local gas temperature but on the temperature of the star $T_*$. We thus set the opacity in these cells to a constant value of $\kappa = 3.5\:\mathrm{cm^2/g}$ which is consistent with $T_*=5780$~K, i.e., a sun-like star \citep{Bitsch14}. }\newline
Our model disk has a gas radial surface density profile $\Sigma_g$ which follows a power law of the form 
\begin{equation}
   \Sigma_g(r) = 80\:\mathrm{g/cm^2}\cdot\Big(\frac{r}{1\:\mathrm{AU}}\Big)^{-1/2} 
\end{equation}
Initially, the dust follows an identical surface density profile but is scaled by a factor $0.01$. {Like in \cite{Binkert21}, the surface density profile corresponds to a total dust mass of $\sim5\cdot10^{-4}\:M_{\odot}$ within 120 au, which is comparable to the most massive disks in the ALMA survey of \cite{Ansdell2016}.} In our models, the mm-sized particles experience Stokes numbers in the range $9\cdot10^{-3} <St <7\cdot10^{-2}$ in the disk midplane (before the insertion of the planet). \newline
Throughout our simulations, we keep the value of the kinematic viscosity constant at a value which corresponds to $\nu=10^{-5}r_0^2\Omega_0$ at a reference radius of $r_0=50$ au. This value corresponds to a Shakura \& Sunyaev $\alpha$-parameter of $\alpha=4.0\cdot10^{-3}$ at $r=50$ au, assuming a vertically isothermal disk with an aspect ratio of $H=0.05$. We purposely set the viscosity at this relatively large value to isolate the effects of planetary stirring and suppress other sources of resolved hydrodynamic turbulence {coming from, e.g., Rossby vortices at gap edges \citep[][]{Zhu2014} or the VSI \citep[][]{Flock2017a,Lin19}}, which, unavoidably, would impact our simulations at lower prescribed viscosity. Therefore, when studying the impact of different strengths of turbulent diffusion, we solely change the diffusion coefficient $D$ and keep the gas viscosity $\nu$ constant to isolate the influence of turbulent diffusion. Throughout this work, we describe the ratio of the used parameters $\nu$ and $D$ with the dimensionless Schmidt number $Sc$ as defined in equation \ref{eq:Sc_number}. However, we stress that in reality, it is the underlying turbulent viscosity that changes and governs the strength of turbulent diffusion and that the Schmidt is expected to always be on the order of unity \citep[][]{Cuzzi1993}. \newline
Like in \cite{Binkert21}, we assume the central star in our simulations to emit a blackbody spectrum with a solar effective temperature $T_\mathrm{eff}=5780\:K$ and have a mass and radius equal to the solar mass and solar radius, respectively ($M_*=M_\odot$, $R_*=R_\odot$). We solve the hydrodynamic equations in spherical coordinates in a rotating frame of reference. Because we are interested in the vertical disk structure, we mainly focus on the radially most extended disk domain presented in \cite{Binkert21} where the vertical extent of the disk is the largest. This domain covers the radial domain between 20 au and 119 au from the central star, with a planet orbiting on a circular orbit with a fixed radius at $r_p=50$ au. In azimuthal direction, we simulate the full $2\pi$ disk, while in polar direction, we assume mirror symmetry about the midplane and include the domain between the disk midplane and $0.13\:\mathrm{rad}$ above the midplane {(corresponding to about three gas scale heights)}. {We keep the numerical resolution of our base grid identical to the one used in our previous study, i.e. $N_\phi\times N_r\times 2N_\theta=680\times215\times 40$, linearly spaced along all dimensions}. With this resolution, we vertically sample a gas scale height $h_g$ with about eight numerical grid cells. In selected simulations, we locally refine the numerical grid in a comoving region surrounding the planet ($\phi=[-1.1,1.1],\:r = [0.5385,1.4615]\cdot r_p,\:\theta=[\pi/2,\pi/2-0.116]$) doubling the resolution along each dimension, i.e., locally increasing the number of cells by a factor eight. We summarize our simulation parameters in \autoref{table:parameters}. In this paper, we are ultimately interested in observational features obtained from synthetic ALMA continuum observations, which are resolution limited by the size of the beam which has a size of about 35 mas (see section \ref{sec:syn_obs}). At 50 au from the central star, the vertical extent of a single grid cell of the base grid in our simulations is equal to 0.32 au, which subtends an angle of $\sim 3.2$ mas at a source-observer distance of 100 pc. Thus, our vertical numerical resolution of the base grid samples the beam about eleven times along one axis. The refined grid samples the beam about 22 times. \newline
In \autoref{table:simulationsets}, we compile a list of all the simulations that we ran for this study. As mentioned, we mainly focus on the \texttt{50au}-domain, for which we designate the simulation containing a Jupiter-mass planet our \textit{fiducial} simulation. We ran this configuration three times with different values of the Schmidt numbers ($Sc=1,10,100$), i.e., changing the value of the diffusion coefficient $D$, to study the influence of turbulent diffusion. We added a grid refinement patch to all three of these simulations. In addition to that, we ran the three identical simulations, but without an embedded planet. Furthermore, we ran simulations containing a more massive 5 Jupiter-mass planet and a less massive Saturn-mass planet with $Sc=1$ in the \texttt{50au}-domain. For comparison, we also ran the Jupiter-mass and 5 Jupiter-mass planet in the \texttt{5au}-domain with $Sc=1$. 
\begin{table}
\begin{center}
\caption{Overview of the physical disk parameters and the of the numerical grid parameter used in the three-dimensional radiative hydrodynamic simulations.}
\begin{tabular}{l l} 
 \hline
 \hline
  Gas surface density & $\Sigma_g = 80 \mathrm{g/cm^2}\cdot \Big(\frac{r}{1\:\mathrm{AU}}\Big)^{-1/2}$     \\ 
  Global dust-to-gas ratio & 0.01  \\
  Stellar parameters & $\mathrm{T_{eff,*}} = 5780\:K$, $\mathrm{M_{*}}=1\:\mathrm{M_\odot}$  \\ 
                     & $\mathrm{R_{*}}=1\:\mathrm{R_\odot}$ \\ 
  \\
  
    Base grid resolution & $N_\phi\times N_r\times 2N_\theta=680\times215\times 40$ \\
  $\phi$-domain      & 0 - $2\pi$ rad \\
  $\theta$-domain  & $\pi/2-0.129$ rad\\
  r-domain "\texttt{50au}" & 20.0 au - 119 au\\  
  r-domain "\texttt{5au}"  & 2.08 au - 12.4 au\\

 \hline
 \hline
\end{tabular}
\label{table:parameters}
\end{center}
\end{table}

\begin{table}
\begin{center}
\caption{List of hydrodynamical simulations conducted in this study.}
\begin{tabular}{l c c c c} 
 \hline
 \hline
 Model \# & r-Domain & Planet Mass $M_p$ & $Sc$ & grid refinement\\
 \hline
    1 & \texttt{50au} & 1 $M_\mathrm{Jup}$ & 1 & $\checkmark$ \\
    2 & \texttt{50au} & 1 $M_\mathrm{Jup}$ & 10 & $\checkmark$\\
    3 & \texttt{50au} & 1 $M_\mathrm{Jup}$ & 100 & $\checkmark$\\
    4 & \texttt{50au} & no planet & 1 & $\times$ \\
    5 & \texttt{50au} & no planet & 10 & $\times$\\
    6 & \texttt{50au} & no planet & 100 & $\times$\\
    7 & \texttt{50au} & 5 $M_\mathrm{Jup}$ & 1 & $\times$\\
    8 & \texttt{50au} & 1 $M_\mathrm{Sat}$ & 1 & $\times$\\
    9 & \texttt{5au} & 5 $M_\mathrm{Jup}$ & 1 & $\times$\\
    10 & \texttt{5au} & 1 $M_\mathrm{Jup}$ & 1 & $\times$\\
 \hline
 \hline
\end{tabular}
\label{table:simulationsets}
\end{center}
\end{table}

\subsubsection{Initial/boundary conditions and simulation procedure}
\label{sec:procedure}
We initialize the gas disk with a constant aspect ratio, $H=h_g/r=0.05$ which then evolves depending on the local heating/cooling when the simulations progress. We initialize the mm-sized dust with a vertical profile equal to the vertical equilibrium distribution in equation (\ref{eq:vertical_dust_solution}). At each radius, the vertically integrated dust-to-gas ratio, i.e., the surface density ratio, initially is equal to $0.01$. During the simulation, the distribution in the gas as well as the dust evolves according to the local thermohydrodynamic conditions and the local dust-to-gas ratio evolves accordingly. Before injecting the planetary gravitational potential, we first run the simulation with only 2 cells in the azimuthal direction for a duration equivalent to 150 planetary orbits to reach the thermodynamic equilibrium of the circumstellar disk. {We set the end of this run as $t=0$.} {Then, we split the azimuthal domain into 680 cells and run the simulation up to $ t=200\cdot2\pi\Omega_0^{-1}$, 200 additional planetary orbits, while we increase the planet's potential to the desired value over the first 100 orbits as in} \cite{Szulagyi2016}. In the relevant simulations (see \autoref{table:simulationsets}), we add the grid refinement patch at $t=200\cdot2\pi\Omega_0^{-1}$ and run the simulation with the locally increased resolution until $t=220\cdot2\pi\Omega_0^{-1}$. \newline
The boundary conditions in the gas fluid are identical to \cite{Szulagyi2016} and \cite{Binkert21}. {Particularly, this means symmetric boundary conditions in the vertical direction, where also the gas density is exponentially tapered based on the local gas temperature at the boundary opposite to the midplane.} In \cite{Binkert21}, we imposed antisymmetric radial boundaries for the radial dust velocity component, which ensured mass conservation in the entire simulation domain. However, at the same time, it allowed dust accumulations to form at the radial boundaries. When including turbulent diffusion, these dust accumulations would diffuse back into the simulation domain. To prevent this, in this work, we also chose symmetric boundary conditions for the radial velocity component, identical to the boundary conditions of other two velocity components and the density. As a result, 3-8 per cent of dust present in the domain is lost via the boundary during the entire evolution of our simulations {(the degree of mass lost scales with the mass of the planet)}.

\subsection{Radiative Transfer}
\label{sec:Radiative Transfer}
After the hydrodynamic simulations, but before post-processing, we exponentially taper the dust density in the inner disk within a region $r<0.5\cdot r_p$, to decrease the impact of potential dust accumulations at the inner computational boundary and/or artifacts caused by the inner computational boundary. \newline
{The need for tapering arises because we do not have density-damping zones in our radiative hydrodynamic simulations. We find dust accumulations at the inner disk edge to build up during the simulations as a result of the incident stellar radiation and the consequent large disk temperature close to the inner disk edge. This is a result of the third term on the r.h.s. of equation (\ref{eq:momentum_conservation_04}). These accumulations resulted in an increased optical depth that affects the radiative transfer post-processing. Thus, we followed the approach of \citet{Speedie22} and truncate the disk in between hydrodynamic simulations and radiative transfer post-processing. We experimented with different values of the truncation radius and found a value of $0.5\cdot r_p$ to be the optimal tradeoff between decreasing the impact that these dust accumulations have on the synthetic observations, and not interfering with the domain of scientific interest.\newline}
We then post-process our set of hydrodynamic simulations with the Monte-Carlo radiative transfer code package \textsc{radmc-3d} \citep{Dullemond2012} in order to create synthetic ALMA observations in band 7 at 0.87 mm. The opacity provided to code is based on a dust grain size distribution of 0.1 $\micron$ and 1 cm with a power-law index of 3.5 \citep{Pohl17} assuming a mixture of silicate \citep{Draine03} and carbon \citep{Zubko1996} with a fractional abundance of 70 per cent and 30 per cent, respectively. The absorption and scattering opacities as well as the g parameter of anisotropy were calculated using the \textsc{bhmie} code of \cite{Bohren1984}. The Bruggeman mixing formula was used to determine the opacity of the mixture. The resulting absorption and scattering opacity at the wavelength of 0.87 mm are $\kappa_\mathrm{abs}$=10.1 cm$^2$/g and $\kappa_\mathrm{sca}$=10.2 cm$^2$/g, respectively. This is comparable to the values in \cite{Birnstiel2018}. Based on these opacities and the dust surface densities, we expect the disk models to be marginally optically thick at the observed wavelength. \newline
The observed mm-continuum fluxes ultimately depend on the distribution of the dust density $\rho_d$ and temperature $T_d$. However, unlike the gas temperature, the dust temperature is not directly calculated in the hydrodynamic simulations. \newline
{In low-density disk regions, e.g., in the disk atmosphere, we expect nonradiative heating/cooling effects to be small and thus the dust temperature to be close to the radiative equilibrium temperature. Complex thermochemical models confirm that this is at least a valid first-order assumption, as shown by e.g., \cite{Woitke22} using {\sc ProDiMo} models \citep{Woitke09}. Models capturing the full disk thermochemistry are desirable but beyond the scope of our current work.} In our work, we find the radiative equilibrium temperature using the thermal Monte Carlo tool \texttt{mctherm} within \textsc{radmc-3d}. However, even in the outer disk, we find the disk midplane, where we expect the majority of the observed thermal emission to come from, to be too dense and consequently dust-gas collisions to be too frequent, to neglect nonradiative cooling/heating effects. Therefore, we refrain from setting the dust temperature equal to the radiative equilibrium temperature and, instead, set the dust temperature equal to the temperature calculated in the radiative hydrodynamic simulations which also accounts for nonradiative cooling/heating effects. \newline
Based on the dust density and temperature distribution, we generate intensity maps at four different inclinations ($i=$ 0$^\circ$, 60$^\circ$, 80$^\circ$, 90$^\circ$) with the \texttt{image} task including the \texttt{fluxcons} argument to assure flux conservation. We assume anisotropic scattering using the Henyey-Greenstein approximation. The size of the intensity maps is 1000 $\times$ 1000 pixels, and we assumed the disk to be at a distance of 100 pc, about the distance to the closest star-forming region. In the \texttt{50au}-domain, this results in an angular resolution of 2.5 mas per pixel. In a subsequent step, the resulting wavelength-dependent intensity maps are processed to simulate ALMA observations (see section \ref{sec:syn_obs}).

\subsection{Synthetic ALMA Continuum Observations}
\label{sec:syn_obs}
We study the observational signatures of a planet in a turbulent disk by generating synthetic continuum observations from the intensity maps using the Common Astronomy Software Applications package \textsc{casa} \citep{CASA2007} to simulate ALMA observations. Particularly, we create observations in band 7 at a wavelength of 0.87 mm (345 GHz) and use antennae configuration C43-8, which provides sufficient resolution and signal-to-noise behavior to pick up small-scale features on the scale of a few AU. With a maximum baseline of 8.5 km, configuration C43-8 allows for an angular resolution of up to 28 mas, which corresponds to a physical scale of 2.8 au at a distance of 100 pc. However, configuration C43-8 in combination with band 7, only has a maximum recoverable scale (MRS) of 410 mas (41 au), smaller than the planetary orbital radius in our radially most extended simulation (50 au). Because we expect observational features such as, e.g., rings on scales comparable to the planetary orbital radius, we also observe the disk with the more compact configuration C43-5 as recommended by the ALMA proposer's guide. The additional configuration increases the MRS in our synthetic observations to 1.94" (194 au), which covers the most relevant angular scales in our simulations. \newline
We set the integration time in the more extended configuration to six hours, and, following the ALMA proposer's guide, set the integration time in the more compact configuration to 79 minutes. For each antennae configuration, we generate a measurement set (MS) with the \texttt{simobserve} task contained in the CASA software package for which we set the channel bandwidth to 7.4 GHz, add thermal noise with a random number seed of 1745, adopt a value of 0.475 mm precipitable water vapor, and set the ambient temperature to 269 K. We combine the two measurements sets with the \texttt{concat} task, before using the \texttt{simanalyze} task to generate the final combined intensity images by applying Briggs weighting to the visibility data (robust=0.5) and setting the \texttt{clean} threshold to a value of 50 \textmu Jy/beam. 

\section{Turbulent Pressure in the Dust Fluid}
\label{sec:turb_diff_in_dust}
Before presenting the results in section \ref{sec:results}, we summarize and illustrate the properties of our turbulent diffusion model, which models turbulent excitations in the dust fluid with an explicit pressure term in the otherwise pressureless dust fluid (see section  \ref{sec:diffusion_pressure_model}). The mass and velocity equations take the form 
\begin{equation}\tag{\ref{eq:dust_cont_eq2}}
\frac{\partial \rho_d}{\partial t}+\bm{\nabla}\cdot(\rho_d \mathbf{v}_d)=0
\end{equation}
\begin{equation}\label{eq:wo_external_forces}
\frac{\partial \mathbf{v}_d}{\partial t}+\big(\mathbf{v}_d\cdot\bm{\nabla}\big)\mathbf{v}_d=-\nabla\Phi-\frac{1}{\tau_s}(\mathbf{v}_d-\mathbf{v}_g)-c_d^2\nabla \ln \frac{\rho_d}{\rho_g}
\end{equation}
where we have dropped the asterisk above the dust velocity for simplicity. {Note, equation (\ref{eq:wo_external_forces}) is derived from the momentum equation in a stratified background, i.e., equation (\ref{eq:momentum_conservation_04}), by applying the product rule to the time derivative and using the continuity equation (\ref{eq:dust_cont_eq2}) to simplify terms.} The first term on the r.h.s. of equation (\ref{eq:wo_external_forces}) models the gravitational acceleration, the second term the drag interaction with the gas, and the third term is the turbulent pressure. Traditionally, the dynamics of the diffusion flux are neglected and the effect of turbulence is modeled as a diffusive source term in the mass conservation equation \citep[like in e.g.][]{Cuzzi1993,Pavlyuchenkov07,Charnoz11,Dullemond18,Weber20}. The governing dust equations in this \textit{mass diffusion model} are the following: 
\begin{equation}\label{eq:adv_diff_equation}
\frac{\partial \rho_d}{\partial t}+\bm{\nabla}\cdot(\rho_d \mathbf{v}_d)=\bm{\nabla}\bigg[D\:\rho_g\:\bm{\nabla}\:\bigg( \frac{\rho_d}{\rho_g} \bigg)\bigg]
\end{equation}
\begin{equation} \label{eq:classical_velocity_equation3}
\frac{\partial \mathbf{v}_d}{\partial t}+\big(\mathbf{v}_d\cdot\bm{\nabla}\big)\mathbf{v}_d=-\nabla\Phi-\frac{1}{\tau_s}(\mathbf{v}_d-\mathbf{v}_g)
\end{equation}
where the diffusion term in equation (\ref{eq:adv_diff_equation}) appears on the r.h.s, while the velocity equation (\ref{eq:classical_velocity_equation3}) remains pressureless. \newline
A drawback of the system (\ref{eq:adv_diff_equation}) and (\ref{eq:classical_velocity_equation3}) is the fact that it can violate the conservation of {linear and angular} momentum \citep[][]{Weber2018,Tominaga19}. However, especially a proper treatment of {angular} momentum is crucial when studying the spatial distribution of gas and dust in turbulent circumstellar disks. We will discuss the specifics of the disk problem in section \ref{sec:Dust Turbulent Diffusion in a Keplerian Disk} but first discuss the differences between the two approaches in an illustrative one-dimensional example. 

\subsection{Dust Turbulent Diffusion in One Dimension}
\label{secturb_diff_1d}
\begin{figure*}
\includegraphics[width=1.95\columnwidth]{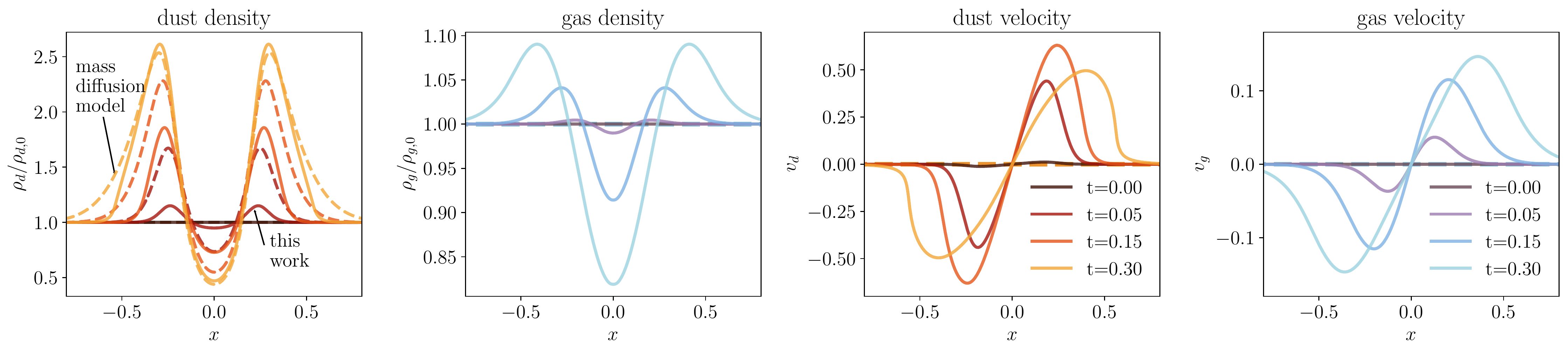}
\caption{Comparison of the one-dimensional diffusive spreading of an initially static Gaussian dust distribution with standard deviation $\sigma = 0.1$ and constant gas background between the mass diffusion model (\textit{dashed lines}), i.e., the solution of equations (\ref{eq:adv_diff_equation}) and (\ref{eq:classical_velocity_equation3}), and the approach taken in this work (\textit{solid lines}), i.e., modeling turbulent dust diffusion as a pressure with equations (\ref{eq:dust_cont_eq2}) and (\ref{eq:wo_external_forces}) (in arbitrary units). {Note, the solid lines in the dust velocity plot represent the effective dust advection velocity $v_d^*$}. The main difference is that in the diffusion pressure approach, the diffusion flux is explicitly associated with a velocity (see third panel). In the mass diffusion model, diffusion does not affect the dust velocity $v_d$. Moreover, the gas is dragged along by the movement of the dust (second and fourth panels). In the first panel, we show the diffusive spreading of the dust density, which is initially slower than in the mass diffusion model because the dust does not react to turbulent fluctuation in the gas on timescales shorter than the stopping time ($\tau_s=0.1$). As a result, the two approaches deviate on small times and on length scales smaller than a characteristic length scale $\lambda_c=2\pi\sqrt{D\tau_s}$ ($\lambda_c\approx0.6$ here). See text section \ref{secturb_diff_1d} for a discussion.}
\label{fig:numtest01}
\end{figure*}

We illustrate the difference between treating turbulent diffusion as a pressure, as we do in this work with equations (\ref{eq:dust_cont_eq2}) and (\ref{eq:wo_external_forces}), and the traditional approach of treating diffusion as pure mass diffusion like in equations (\ref{eq:adv_diff_equation}) and (\ref{eq:classical_velocity_equation3}), in an illustrative example similar to the one presented in \cite{Huang22}. \newline
We set up a static one-dimensional isothermal viscous gas background with a ten per cent dust content by mass ($\rho_d(x)=0.1$, $\rho_g(x)=1.0$) in arbitrary units and in the absence of external forces ($\nabla\Phi=0$). At $t=0$, we add a Gaussian to the dust background centered around $x=0$, with standard deviation $\sigma = 0.1$ and amplitude 0.9 and let it diffusively spread over time. We keep all other relevant parameters constant (sound speed $c_s=1$, gas viscosity $\nu = 0.1$, dust diffusion coefficient $D=0.1$). We also keep the stopping time constant ($\tau_s=0.1$) and show the evolution of the gas and dust densities and velocities in \autoref{fig:numtest01}. We plot the solution of the mass diffusion model obtained by solving equations (\ref{eq:adv_diff_equation}) and (\ref{eq:classical_velocity_equation3}) with dashed lines at four different points in time ($t=0,\:0.05,\:0.1,\:0.2$). The diffusive pressure solution (equations (\ref{eq:dust_cont_eq2}) and (\ref{eq:wo_external_forces})) is plotted with solid lines. The first plot in the \autoref{fig:numtest01} shows the diffusive spreading of the normalized dust density. The mass diffusion model leaves the gas density (second plot), dust velocity (third plot) and gas velocity (fourth plot) at its initial value of zero. On the other hand, the diffusive pressure approach has a velocity associated to the outwardly directed diffusion flow (see third panel of \autoref{fig:numtest01}). As a result of the drag interaction, the diffusive motion of the dust, drags along the gas, moving the gas outward along with the dust (see second and fourth panel of \autoref{fig:numtest01}). This behavior is also reported in \cite{Huang22} who likewise report angular momentum conservation in their diffusive multifluid approach but do not model diffusion as a pressure term. \newline
We predict in appendix \ref{sec:One Dimension in the Absence of External Forces} by means of a linear perturbation analysis of the relevant equations, that the dust does not couple to the turbulent motions of the gas on timescales shorter than the stopping time. Therefore, the pressure-driven diffusive spreading is initially slower than in the mass diffusion model and takes until $t\sim0.2$ until it has caught up (see the first plot of \autoref{fig:numtest01}). In appendix \ref{sec:One Dimension in the Absence of External Forces}, we also show that the dust does not fully couple to turbulence fluctuations on length scales smaller than a characteristic length scale $\lambda_c=2\pi\sqrt{D\tau_s}$. In the current example, this corresponds to a value of $\lambda_c\approx0.6$. Indeed, in the first panel of \autoref{fig:numtest01}, we find the peaks of the dashed and the solid curves to only merge once they approach $x\sim0.6$ where about two stopping times have passed. At later times, i.e., on larger spatial scales, the dust densities in the two diffusion approaches align but do not exactly match because, in the diffusion pressure approach, the gas reacts to the spreading of the dust (see the second panel of \autoref{fig:numtest01}) which again feeds back on the distribution of the dust. 

\subsection{Dust Turbulent Diffusion in a Keplerian Disk}
\label{sec:Dust Turbulent Diffusion in a Keplerian Disk}
After briefly discussing our model in the absence of external forces in section \ref{secturb_diff_1d}, we now include gravitational forces in the specific application of a (non-self-gravitating) rotating protoplanetary disk. The main standard that we compare our model to, is the seminal work of \cite{Youdin2007} who studied the effects of particle stirring in turbulent gas in a protoplanetary disk environment in great detail.

\subsubsection{Radial Turbulent Diffusion}
\label{sec:rad_turb_diff}
\cite{Youdin2007} showed that epicyclic oscillations of moderately to weakly coupled dust grains in the $r-\phi$-plane of a protoplanetary disk, weaken the effects of turbulent diffusion in the radial direction such that the effective radial diffusivity in a Keplerian flow scales with the Stokes number as $1/(1+St^2)$. Because this effect is not captured in the mass diffusion model, it is usually explicitly prescribed by scaling the diffusion coefficient in the radial direction with this additional factor based on the local Stokes number $St$, such that $D_\mathrm{eff}=D/(1+St^2)$ \citep[e.g.][]{Dullemond18}. \newline
In this work, we do not explicitly prescribe the factor $1/(1+St^2)$ but show in appendix \ref{sec:Two-Dimensional Diffusion in a Keplerian Disk} by means of a linear perturbation analysis that, as a result of {angular} momentum conservation in our model, the effective radial diffusivity implicitly decreases for larger Stokes numbers exactly as $D_\mathrm{eff}=D/(1+St^2)$ (see \autoref{fig:D_vs_St}). Moreover, in section \ref{sec:Radial Turbulent Diffusion in an Axisymmetric Disk}, we also confirm this result numerically by studying the diffusive radial spreading of an axisymmetric dust ring in a protoplanetary disk and find the ring to diffuse less efficiently with increasing Stokes number when angular momentum is conserved (see \autoref{fig:2D_diffusion_test}). In conclusion, we find that by conserving {angular} momentum, we implicitly capture the effects of epicyclic oscillation of moderately and weakly coupled grains on turbulent diffusion. Furthermore, the implicit nature of our approach has the great advantage that it also captures the effects of epicyclic oscillations if the flow structure in the protoplanetary disk deviates from a purely Keplerian flow and the epicyclic frequency locally varies, e.g., in the surroundings of an embedded planet, a regime in which an explicit approach is not as straightforward to prescribe. 

\subsubsection{Vertical Settling-Diffusion Equilibrium}
\label{sec:Vertical Settling-Diffusion Equilibrium}
Besides diffusion in the radial direction, we also compare the vertical equilibrium dust structure in a protoplanetary disk to the detailed results of \cite{Youdin2007}. Assuming the eddy time is at most equal to the inverse of the Keplerian frequency $\Omega^{-1}$, they find
$h_d^2=\delta/(St+\delta)h_g^2$. Where $h_d$ is the dust scale height and $h_g$ is the gas scale height. The parameter $\delta$ is the dimensionless diffusivity, which is defined such that 
\begin{equation}\label{eq:definition_dimless_delta}
    D = \delta c_s h_g
\end{equation}
In the mass diffusion model, one typically uses the terminal velocity approximation to find the solution that represents the vertical density distribution in the vertical settling diffusion equilibrium \citep[e.g.][]{Schrapler04,Fromang2009}. However, the terminal velocity approximation is only valid for $St\ll 1$, and thus, the formal solution is strictly not valid when the Stokes number approaches unity \citep[][]{Youdin2007,Hersant2009}. \newline
In this work, we find the vertical equilibrium distribution without the need for assuming terminal velocity, but via the force balance in equation (\ref{eq:momentum_conservation_04}), in which the vertical acceleration due to diffusion exactly balances gravity. We approximate the vertical component of the gravitational acceleration with $\nabla_z\Phi=-\Omega^2 z$, which holds for geometrically thin disks. Assuming a static solution (i.e., time derivatives and velocities vanish), and assuming the gas background follows a Gaussian profile with scale height $h_g$ (i.e., it is vertically isothermal), we first derive the vertical dust profile for a constant stopping time ($\tau_s=\mathrm{const.}$). Then, equation (\ref{eq:momentum_conservation_04}) in vertical direction simplifies to
\begin{equation}\label{eq:vertical_balance}
\frac{1}{\rho_d}\frac{\partial}{\partial z}\Big(\rho_d\Big)=-\bigg(\frac{1}{c_d^2}+\frac{1}{c_s^2}\bigg)\Omega^2z.
\end{equation}
which can easily be integrated to give a solution of the form 
\begin{equation}\label{eq:vertical_dust_solution}
    \rho_d(z)=\rho_{d,0}\exp\bigg( -\frac{z^2}{2h_d^2}\bigg)
\end{equation}
where we have defined the dust scale height $h_d$ such that 
\begin{equation}
    h_d^2=\frac{c_d^2}{c_s^2+c_d^2}h_g^2
\end{equation}
which is equivalent to
\begin{equation}\label{scale_height_ratio}
    h_d^2=\frac{\delta}{St+\delta}h_g^2
\end{equation}
and is in agreement with the results of \cite{Youdin2007}.\newline
Next, we again consider the vertical force balance and equation (\ref{eq:vertical_balance}), but this time, we allow the stopping time $\tau_s$ to vary vertically according to its definition in equation (\ref{eq:stopping_time}). The integration of equation (\ref{eq:vertical_balance}) gives the following vertical dust density profile: 
\begin{equation}\label{eq:vertical_equilibrium_profile}
    \rho_d(z)=\rho_{d,0}\exp\bigg[ -\frac{St_\mathrm{mid}}{\delta}\bigg( \exp\bigg(\frac{z^2}{2h_g^2} \bigg)-1\bigg)-\frac{z^2}{2h_g^2}\bigg]
\end{equation}
where $St_\mathrm{mid}$ is the Stokes number evaluated at the midplane. We note that equation (\ref{eq:vertical_equilibrium_profile}) is identical to the equilibrium profile found in \cite{Fromang2009} in their equation (19), but without using the terminal velocity approximation. Thus, in our model, the validity of equation (\ref{eq:vertical_equilibrium_profile}) is formally extended to the weakly coupled regime ($St>1$).

\subsubsection{Numerical Considerations}
Besides the advantages of conserving angular momentum, implicitly capturing the in-plane effects of epicyclic oscillations, correctly modeling the vertical equilibrium distribution across all coupling regimes, and fulfilling the good mixing condition, our diffusion model as introduced in section \ref{sec:diffusion_pressure_model} also has some advantages when it comes to numerical implementations. 
In the mass diffusion model, dust is usually modeled as a pressureless fluid and the diffusion term in the continuity equation (\ref{eq:adv_diff_equation}) is generally solved with a finite differences approach on a grid, either directly in the numerical advection step or an operator splitting approach is used to solve the diffusion term in a separate source step. Either way, difficulties may arise when computing spatial derivatives when the dust density distribution exhibits small-scale structures down to the scale of the numerical grid. We found finite differences methods to be numerically unstable around such small-scale, poorly resolved density structures, leading to zigzag-shaped dust density features, especially in three-dimensional stratified simulations. As a countermeasure, one could resort to the incorporation of artificial viscosity to stabilize the numerical scheme \citep[like it was done in e.g.][]{Zhu2012}, but we find this to be an unfavorable solution. On the other hand, numerically solving equations (\ref{eq:dust_cont_eq2}) and (\ref{eq:momentum_conservation_03}) (or equation (\ref{eq:momentum_conservation_04}) respectively), which include the pressure-like turbulent diffusion, has the advantage that there is no explicit diffusion term in the continuity equation because the diffusion terms are all contained in the  divergence term of the momentum equation (\ref{eq:momentum_conservation_03}). By applying the divergence theorem on the numerical grid, we remove the spatial derivatives on the l.h.s of equation (\ref{eq:momentum_conservation_03}). Then, the only task left is to evaluate cell interface fluxes via Riemann solvers, an approach that is more numerically stable. Thus, the numerical scheme to solve the dust equations is identical to the scheme used to solve the (isothermal) gas equations and there is no need for an additional pressureless fluid solver \citep[see e.g.][]{LeVeque2004}.  

\section{Results}
\label{sec:results}
In this section, we present our results by first qualitatively discussing the three-dimensional dust morphology in the presence of turbulent diffusion and an embedded planet in section \ref{sec:Three-dimensional Dust Morphology}. There, we mainly focus on the simulation containing a Jupiter-mass planet orbiting on a circular orbit at 50 au from the central star and Schmidt number $Sc=100$. {This particular example was chosen because the flow pattern due to planetary mixing is best identified/studied in a background with weak turbulent stirring.} In section \ref{ref:Effective Diffusivity}, we then quantify the level of planetary dust mixing with an effective diffusivity before presenting the synthetic ALMA observations in section \ref{sec:Synthetic Continuum Observations}.

\subsection{Three-dimensional Dust Morphology}
\label{sec:Three-dimensional Dust Morphology}
\begin{figure}
\includegraphics[width=0.95\columnwidth]{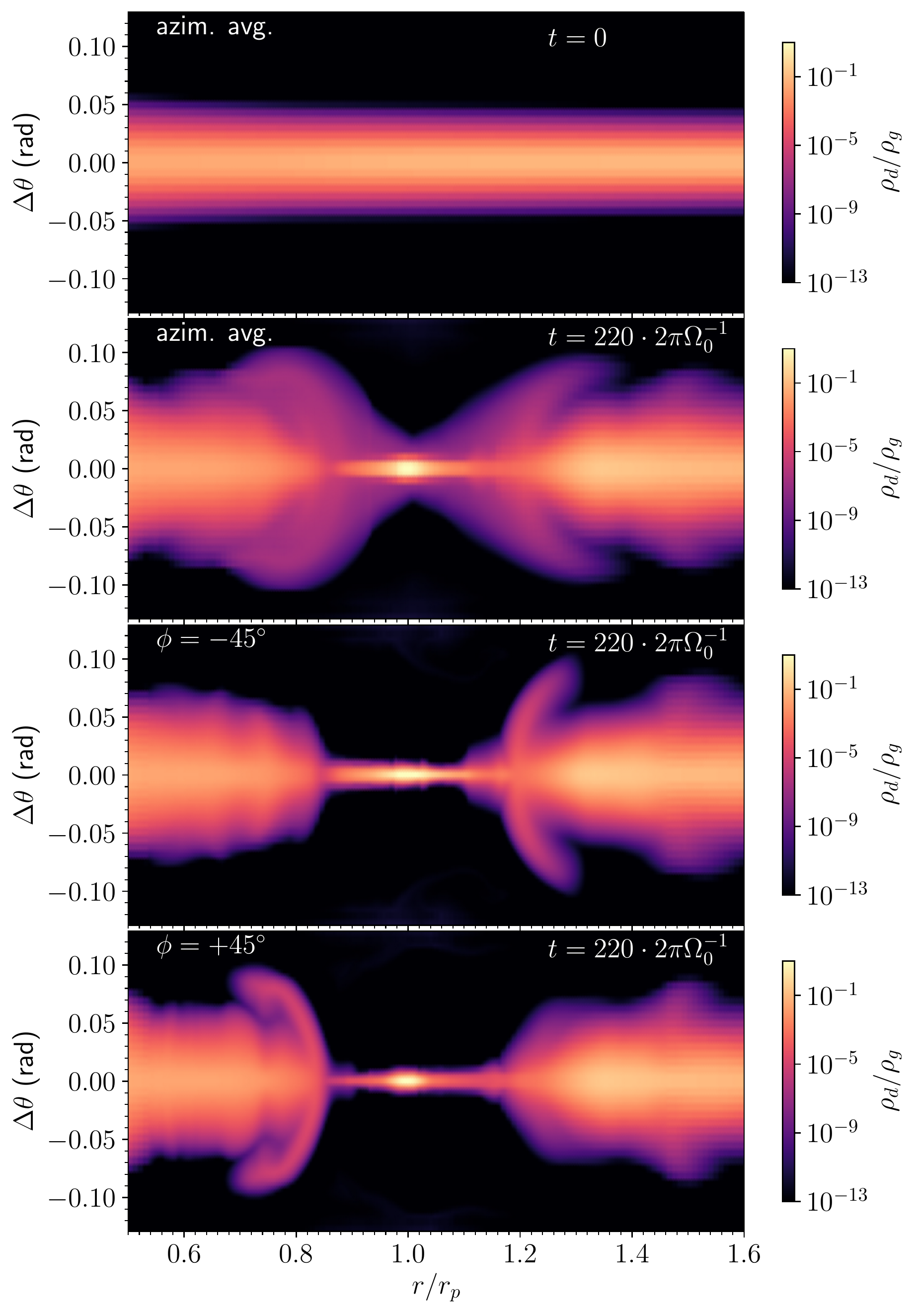}
\caption{Dust-to-gas ratio in our fiducial simulation, containing a Jupiter-mass planet orbiting at $r_p=50$ au from the central star with reduced dust diffusivity compared to the fiducial model ($Sc=100$). The first subplot shows the azimuthally averaged dust-to-gas ratio at $t=0$, i.e., before injecting the planet potential. The second subplot shows the azimuthally averaged dust-to-gas ratio at $t=220\cdot 2\pi\Omega_0^{-1}$. The third and fourth subplots show a vertical cut at $\phi=-45^\circ$ and $\phi=+45^\circ$. The gap region and the planetary stirring of dust can clearly be identified. }
\label{fig:dust_vertical_Sc100}
\end{figure}

In the first sub-panel of \autoref{fig:dust_vertical_Sc100}, we show the azimuthally averaged dust-to-gas ratio of the \texttt{50au} simulation with $Sc=100$ at time-zero (before inserting the planet). Turbulent diffusion counteracts the vertical settling of the mm-sized dust grains such that, in the absence of resolved turbulent flows and/or additional gravitational forcing, the vertical disk profile does settle in an equilibrium distribution in which downward vertical settling is perfectly balanced by upward turbulent diffusion (see section \ref{sec:Vertical Settling-Diffusion Equilibrium}). With the term \textit{downward}, we refer to the direction towards the disk midplane and with \textit{upward} the direction away from the disk midplane. Note that the gas in our radiative simulations is not necessarily vertically isothermal and therefore, the vertical profile of the dust-to-gas ratio does not necessarily follow the equilibrium profile of equation (\ref{eq:vertical_equilibrium_profile}). \newline
We slowly introduce the planetary potential to the circumstellar disk. The embedded planet, modeled via its gravitational potential, then becomes a source of additional dust mixing besides the background level of turbulent diffusion \citep[][]{Binkert21}. In the second sub-panel of \autoref{fig:dust_vertical_Sc100}, we show the azimuthally averaged dust-to-gas ratio of the simulation containing a Jupiter-mass planet at 220 orbits (t = 220 $2\pi\Omega_{p}^{-1})$ in the \texttt{50au} with $Sc=100$. The third and fourth sub-panels show vertical cuts at $\phi=\pm45^\circ$ (the planet is located at $\phi=0^\circ$). The vertical distribution shows the characteristic vertical plume-like structures to the inside and outside the planetary orbital radius at $r=r_p$ that are a result of dust stirring caused by meridional flows \citep[][]{Szulagyi2022}. Planetary dust stirring in the absence of background turbulent diffusion was previously reported in \cite{Binkert21} and \cite{Bi2021} and further confirmed by \cite{Krapp22} in vertically isothermal simulations including turbulent diffusion. As opposed to \cite{Bi2021}, who report the vertically puffed up dust distribution to be roughly azimuthally symmetric, we find an asymmetric distribution with respect to the planet, which becomes apparent when comparing the third and fourth sub-panel of \autoref{fig:dust_vertical_Sc100}. We find the dust distribution to be azimuthally more symmetric in simulations containing a less massive planet and/or more strongly coupled dust. \newline
Vertical flows in the gas, as part of a meridional circulation, have previously been found in hydrodynamic simulations \citep[][]{Kley2001,Szulagyi2014,Fung2016} and have also been confirmed observationally \citep[][]{Teague19}. The existence of similar flow structures in the solid disk component \citep[][]{Szulagyi2022} could have relevant consequences on, e.g., dust grain chemistry as grains experience different chemical and physical environments, or it is relevant for grain growth when flow structures influence the relative velocities of individual dust grains. Further, large-scale dust flows could influence the three-dimensional disk morphology and thus have observational consequences for continuum emissions, which directly trace the spacial distribution of dust grains. To investigate potential observational impacts of planet-induced dust stirring, we further analyze the origin and spatial structure of the vertical dust features in the remainder of this section. \newline
\begin{figure*}
\includegraphics[width=1.95\columnwidth]{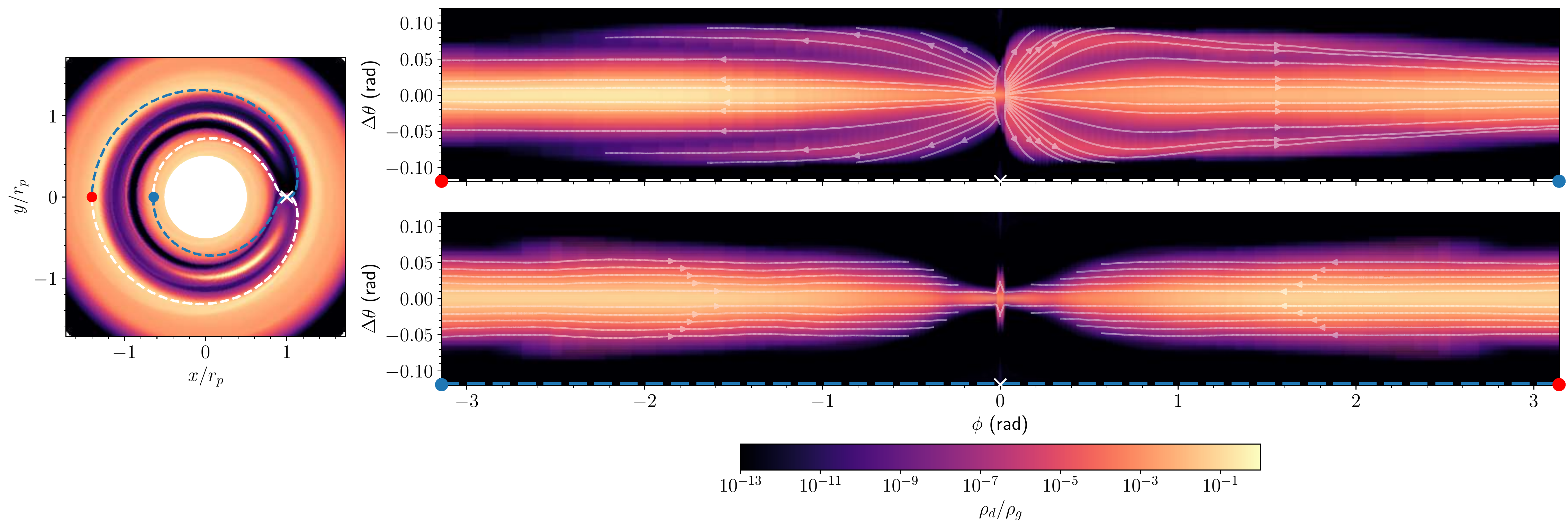}
\caption{Visualization of how dust is delivered to the planetary region by the Keplerian flow (bottom right plot), where it is lifted and then transported away from the planet on the opposing side (upper right plot). \textit{Left:} Dust surface density map of the simulation containing a Jupiter-mass planet orbiting at 50 au. The dashed blue and white lines indicate the curves along which we plot the vertical cuts on the r.h.s. \textit{Right:} Vertical cuts along the blue (\textit{top}) and white (\textit{bottom}) curves, which each cover the full azimuthal range of $\pm\pi$. The color map represents the vertical distribution of the local dust-to-gas ratio $\rho_d/\rho_g$ along these curves, with $\phi=0$ being the position of the planet and positive in the counterclockwise direction. The {faint white} streamlines {visualize} the dust velocity components in the polar direction and parallel to the curve in the co-rotation frame of the planet. Note that these streamlines represent the velocity field along a two-dimensional surface and are not fully representative of the three-dimensional flow. For better visualization, the vertical axes of the plots on the r.h.s. are stretched by a factor $\sim 4.5$.}
\label{fig:vertical_streamlines}
\end{figure*}
%######################################################################
We still focus on the simulation containing the Jupiter-mass planet with $Sc=100$ and examine the dust flow structure there, before generalizing our results to different sets of parameters. {We find distinct flow structures in the mm-sized dust, which are created by the planet and are inherently three-dimensional, and vary strongly in space.} Thus, it is difficult to visualize them in two-dimensional plots. We especially found cuts in the r-z-plane or azimuthal averages of density distributions or velocity fields, e.g., like in \autoref{fig:dust_vertical_Sc100}, to poorly represent the underlying nature of the dust distribution and flow structure. In order to improve upon previous explanations and visualizations of planet-induced dust stirring, we show vertical cuts along a specific curve (empirically determined) in $r-\phi$ space in \autoref{fig:vertical_streamlines}. Specifically, we show the vertical dust density distribution along two curves with the functional dependency
\begin{equation}\label{eq:trajectory:_formula}
    \frac{r_i(\phi)}{r_p} = sgn(\phi)A_i\abs{\phi}^b_i+1
\end{equation}
where $\phi \in (-\pi,\pi)$. For the first curve, we chose $A_1=-0.24$ and $b_1=0.35$ for $\phi\geq0$ and $A_1=-0.27$ and $b_1=0.35$ for $\phi\leq0$. For the second curve, we chose $r_2(\phi) = r_1(-\phi)$. These two curves are represented by a white and blue dashed line respectively in the surface density plot on the l.h.s of \autoref{fig:vertical_streamlines}. The r.h.s plots in \autoref{fig:vertical_streamlines} represent the vertical cuts along these two curves and show the dust-to-gas ratio. Moreover, we show the streamlines of the vertical and parallel dust velocity components along the two curves in the co-rotating frame of the planet. Note that these streamlines only represent the velocity field along a two-dimensional surface and do not fully represent the three-dimensional flow. However, they nicely visualize the influence of the planet on the dust flow. \newline
{Note that equation (\ref{eq:trajectory:_formula}) is functionally similar to the spiral wake parametrization of \citet{Rafikov02}, who predict the functional form of planet-generated density waves in a gas disk. For a disk with $c_s\propto r^{-1/4}$, equation (44) of \citet{Rafikov02} predicts a wake profile of $r\propto \phi^{0.8}$ (for $\phi >0$) far away from the planet, which is comparable to the profile of the spiral density wakes present in the gas. The profile described by equation (\ref{eq:trajectory:_formula}) is with $r\propto \phi^{0.35}$ more tightly wound and distinctly different from the spiral wake in the gas (see also the discussion in section \ref{ref:Effective Diffusivity}, and compare to the tightly wound flow feature in Figure \ref{fig:2d_map_diffusion_coefficient}).} \newline
Along both, the blue and the white curve, the width of the background distribution narrows as it approaches the location of the planet. In the upper sub-plot on the r.h.s of \autoref{fig:dust_vertical_Sc100}, a wing-like structure is superimposed on the smooth background distribution. Such a feature is absent in the lower subplot. It is these wing-like structures that are responsible for the vertical features seen in \autoref{fig:dust_vertical_Sc100}. We find that the wing-like structures are caused by vertical flows in the dust that are strongest at the location where the planetary spiral wake intersects with the edge of the gap ($\phi\sim\pm 0.4$). The wing-like features are asymmetric with respect to the planet, with the feature associated with the inner gap edge being more extended in the polar direction. \newline
%gas flow
Our simulations are strictly symmetric about the midplane. Therefore, the observed vertical flows are not a direct result of the local gravitational field. Instead, we find the vertical dust flows to be driven by the vertical roll-up motions of the gas in the wake of the planet. These distinct flows in the gas in the presence of a planet were first reported in \cite{Szulagyi2014} and \cite{Fung2015} and are part of the meridional circulation created by the planet. The origin of the vertical upward motion in the gas can be understood by considering the gas flow in the planet's co-rotation frame. In such a frame, gas approaches the planet from two sides on a horseshoe orbit. Away from the planet, this gas is vertically in hydrostatic equilibrium and thus roughly flows with a columnar structure. As the column approaches the planet, it enters the Hill sphere of the planet, where the flow components away from the midplane rapidly accelerate vertically toward the disk midplane because the increased vertical gravity of the planet breaks the vertical hydrostatic equilibrium. Thus, a portion of the gas flow on the horseshoe orbit has lost significant potential energy as it arrives at the turn of the horseshoe (when the flow crosses $r = r_p$ and is closest to the planet), and thus has gained kinetic energy (e.g., see Figure 5 in \cite{Fung2015} for a visualization of the gas flow structure at the horseshoe turn). After the horseshoe turn, the fast-moving gas then radially moves away from the planetary orbital radius close to the midplane. \cite{Fung2015} call this component of gas flow, which is pulled toward the planet from high altitudes and continues radially at midplane, the \textit{transient horseshoe flow}. They call it \textit{transient} because, due to the excess radial speed, the gas flow is no longer part of the recurring horseshoe flow. Instead of following the horseshoe trajectory, the gas flow overshoots and exits the horseshoe region, where it encounters the Keplerian flow that flows along quasi-circular orbits outside the horseshoe region (unless it enters the planet's Bondi sphere where it becomes part of the \textit{atmospheric recycling} flow \citep[e.g.][]{Ormel2015,Kuwahara2019}. The fast-moving radial flow enters the quasi Keplerian flow field exactly where the streamlines of the approaching Keplerian flow are bent toward the planet at Lindblad resonances. The result is a convergence of the two flow components at close to 90 degrees (similar to the description in \cite{Szulagyi2014}), which further increases the local gas pressure at this location. This local non-equilibrium build-up of gas pressure decompresses in an upward direction via the vertical roll-up motion, discussed in \citep[][]{Szulagyi2014,Fung2015,Szulagyi2022}. The forced upward motion at the location of the convergence of the two flow components is also the origin of the upward-directed part of the meridional circulation. In this work, we find that the fast midplane gas flows which are deflected upwards drag along a substantial amount of mm-sized dust causing the characteristic plumes on two opposing sides of the planet, which we visualize in the upper right sub-plot of \autoref{fig:vertical_streamlines}. Note how all the streamlines in this plot originate in a region close to the planet at the midplane. I.e., dust that is lifted to regions above the midplane, mainly originates from a region close to the midplane and the resulting effect is the large-scale vertical planetary dust mixing. The fact that the strong vertical gas flows induced by the embedded planet may also drag along substantial amounts of dust to high-altitude disk regions was already hypothesized by \cite{Edgar2008}. Vertically, the dust plumes extend roughly a Hill radius $\sim r_H$, in agreement with the predicted scale for the gas \citep[][]{Fung2015}. Downstream, the component of the flow (gas and dust) which has been lifted vertically away from the midplane is carried away from the azimuthal location of the planet by the differentially rotating Keplerian disk flow. We find that away from these particular, spatially very localized, upward gas motions, vertical stirring is not sustained, and the vertically lifted dust settles into its vertical equilibrium distribution. If the dust grains are only marginally coupled, as is the case in our fiducial simulation, they completely settle before they encounter the planet again \citep[see also][]{Szulagyi2022}. The result is a strong asymmetry in the distribution of the mm-sized dust along the orbit of the embedded planet. After this qualitative description of the relevant physics in the example shown in \autoref{fig:vertical_streamlines}, we study the planetary dust stirring more quantitatively in the next section. 

\begin{figure*}
\includegraphics[width=0.65\columnwidth]{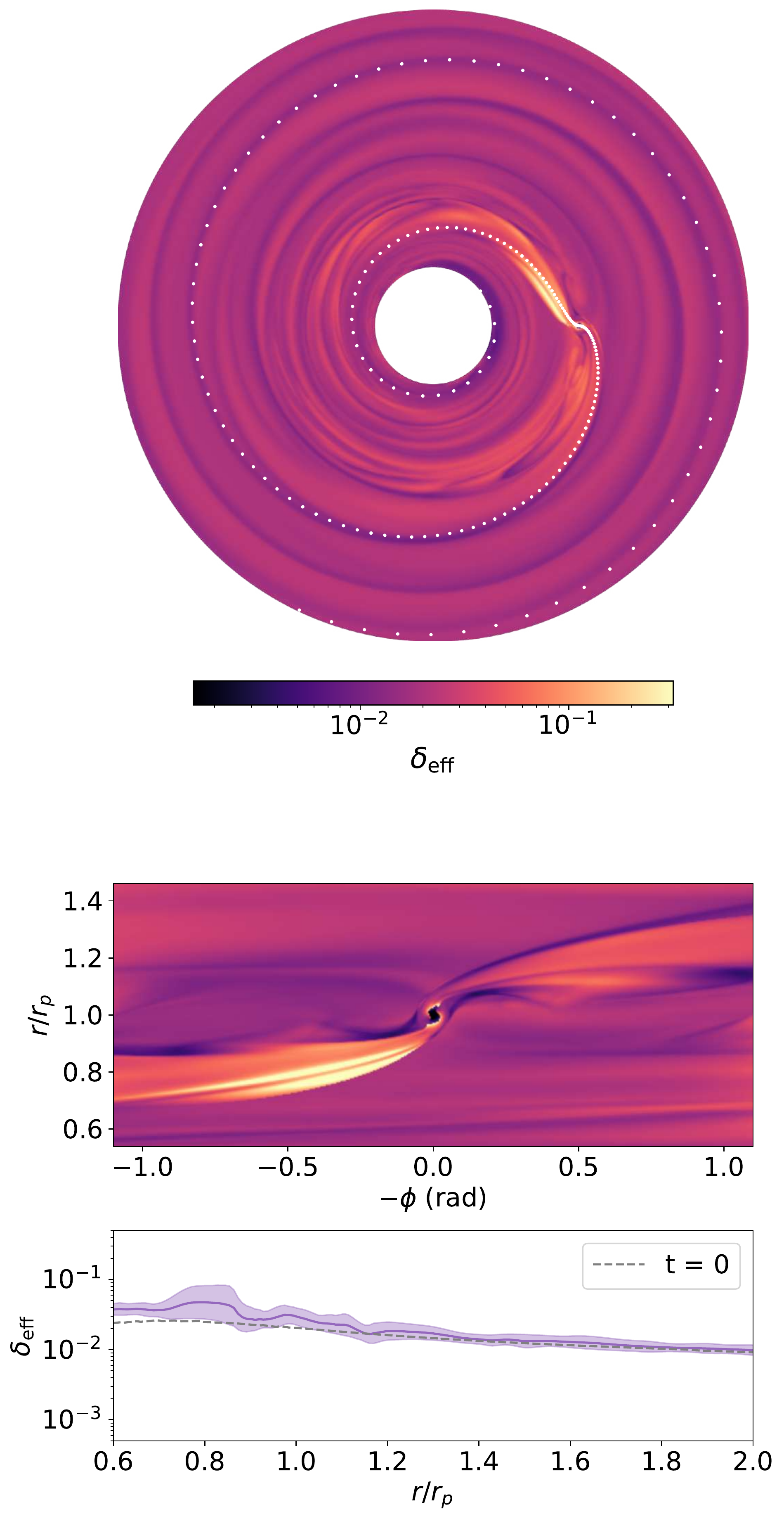}
\includegraphics[width=0.65\columnwidth]{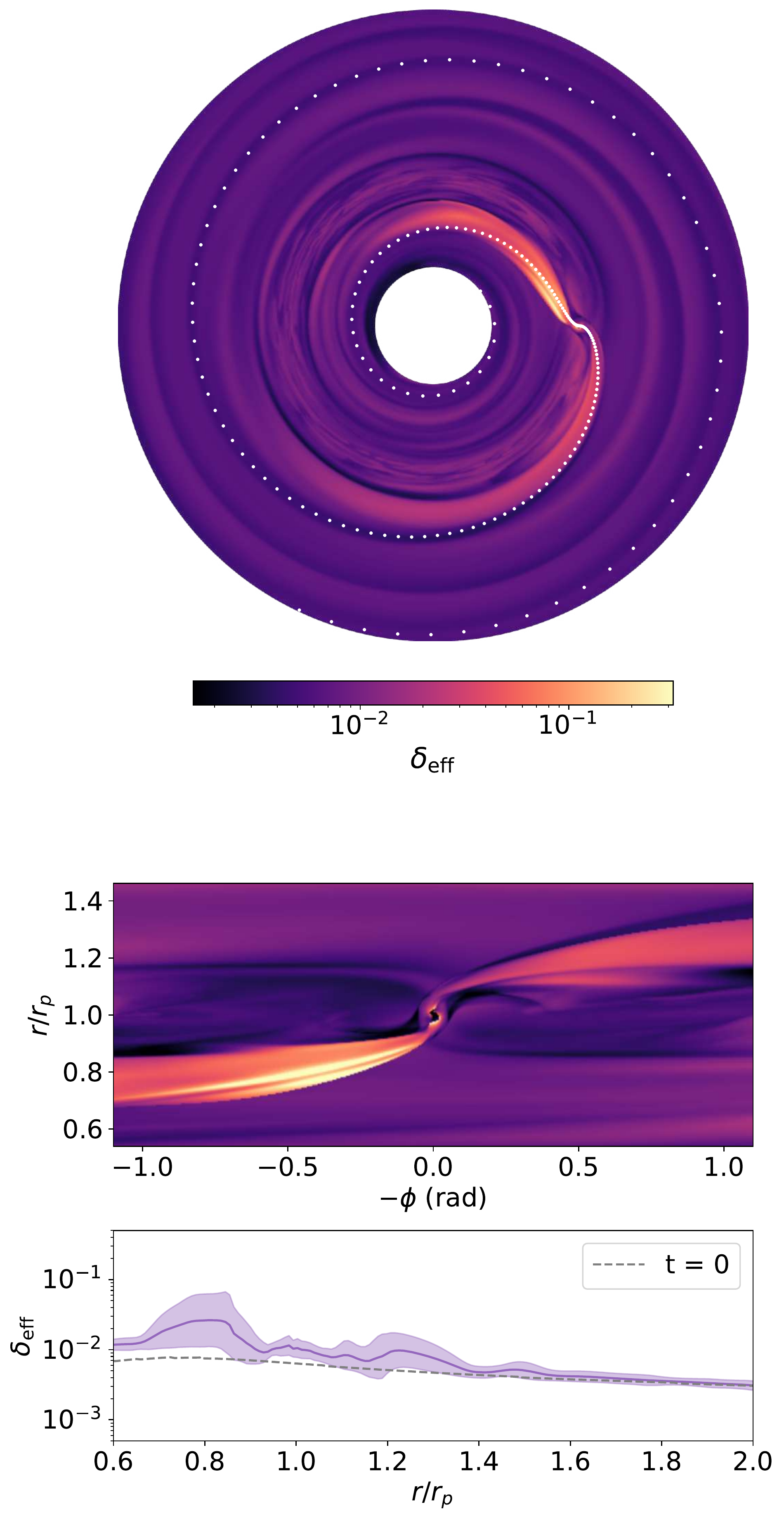}
\includegraphics[width=0.65\columnwidth]{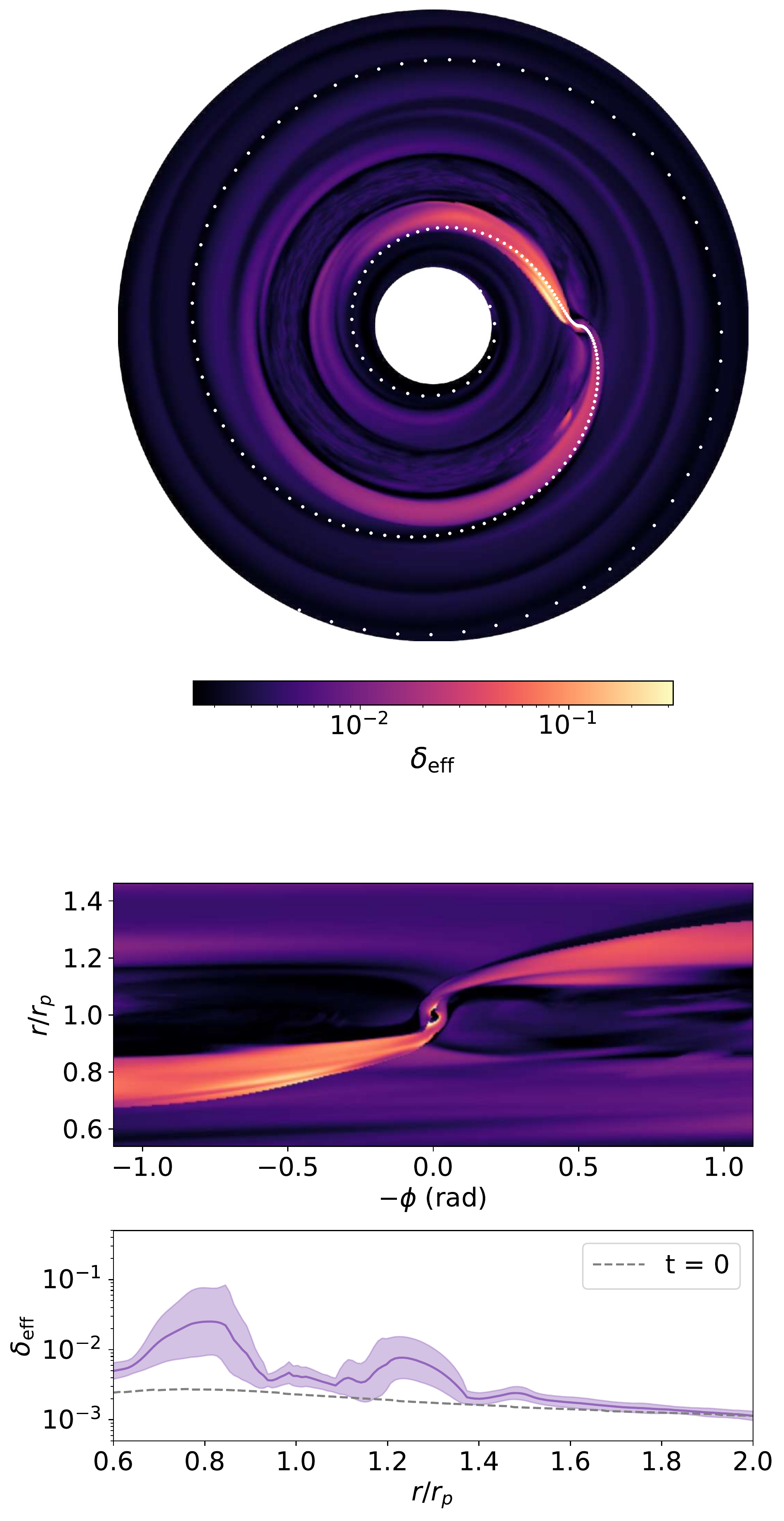}
\caption{This figure shows the effective vertical diffusivity $\delta_\mathrm{eff}$ as found by fitting equation (\ref{eq:vertical_equilibrium_profile}) at every point in the x-y-plane in three simulations of the \texttt{50au}-domain containing a Jupiter-mass planet with decreasing strength of turbulent diffusion from left to right ($Sc=1,10,100$). \textit{Top:} Two-dimensional map of the effective vertical diffusivity $\delta_\mathrm{eff}$. {For comparison, the white dots follow the wake equation of \citet{Rafikov02} and matches well the spiral wake in gas.} The \textit{middle} panels show a zoomed-in view of the region surrounding the planet with increased numerical resolution. \textit{Bottom:} The solid line traces the azimuthal average of the effective vertical diffusivity $\delta_\mathrm{eff}$. The shaded regions show the one sigma deviation from the average value, and the dashed line is the unperturbed average value at $t=0$.}
\label{fig:2d_map_diffusion_coefficient}
\end{figure*}

\subsection{Effective Diffusivity}
\label{ref:Effective Diffusivity}
In the previous section \ref{sec:Three-dimensional Dust Morphology}, we have qualitatively described the vertical dust stirring by a planet embedded in a circumstellar disk. In this section, we aim to quantify the level of vertical planetary stirring and how it is influenced by turbulent diffusion, by measuring an effective diffusivity $\delta_\mathrm{eff}$. \newline
Ultimately, we are interested in how planetary dust stirring is affected by different strengths of dust turbulent diffusion. The straightforward approach to expose the planetary environment to different levels of turbulent diffusion is to change the turbulent viscosity $\nu$ in the gas because when keeping the Schmidt number at unity (see equation \ref{eq:Sc_number}), a change in viscosity will also change the strength of turbulent diffusion. However, as mentioned in section \ref{sec:method1}, we found that lower values of the gas viscosity give rise to additional sources of dust stirring, likely attributed to the VSI and/or vortices at the gap edges generated by the Rossby wave instability. These additional effects make it difficult to isolate and study the sole effects of planetary dust stirring. Therefore, in this study, we keep the gas viscosity fixed at the relatively large fiducial value of $\nu=10^{-5}r_0^2\Omega_0$ in order to suppress additional sources of dust stirring. Nonetheless, we aim to explore the effects of different levels of dust turbulent diffusion and thus alter the value of the dust diffusion coefficient $D$ while keeping the gas viscosity at the fiducial value the same. We thus effectively change the value of the Schmidt number $Sc$ (see equation \ref{eq:Sc_number}). Besides our fiducial setups, we ran additional simulations in which we decrease the dust diffusion coefficient by one and two orders of magnitude, respectively, and leaving the remaining parameters identical (see \autoref{table:simulationsets}). Thus, we ran simulations with different levels of dust turbulent diffusion in which the Schmidt number, as defined in equation \ref{eq:Sc_number}, takes values of $Sc=1,10,100$, such that lower levels of dust turbulent diffusion are associated with a larger Schmidt number.\newline
In section \ref{sec:Vertical Settling-Diffusion Equilibrium} and equation (\ref{eq:vertical_equilibrium_profile}), we showed that, for a given (vertically isothermal) gas distribution, the vertical extent of the dust depends on the ratio between the midplane Stokes number $St_\mathrm{mid}$ and the dimensionless diffusion coefficient $\delta$. In our simulations, we determine the midplane Stokes number $St_\mathrm{mid}$ and the gas scale height $h_g$ at every coordinate ($x,y$) and approximate the local vertical dust density profile $\rho_d(z)$ at this location with equation (\ref{eq:vertical_equilibrium_profile}) and determine an effective diffusivity $\delta_\mathrm{eff}(x,y)$ by doing a least squares fit in log-space. {We stress here that the vertical dust density structure is strictly not in a vertical settling-diffusion equilibrium wherever the flow is highly dynamic, e.g., in the planetary wakes. Nonetheless, our approach allows us to quantify the level of vertical stirring with an effective diffusivity $\delta_\mathrm{eff}$.} \newline
We visualize our results in \autoref{fig:2d_map_diffusion_coefficient} where we show two-dimensional maps of the effective diffusivity $\delta_\mathrm{eff}$ in the upper row of each subplot. The sub-panels show, from left to right, simulations with decreasing levels of turbulent diffusion. While the sub-panels of the left side show the fiducial simulation, the parametrized dust diffusion coefficient is decreased by a factor 10 in the simulation shown in the sub-panels in the middle and decreased by a factor 100 in the sub-panels on the right. All three columns show a simulation containing a Jupiter-mass planet orbiting at 50 au. \newline
The subplots in the second row of \autoref{fig:2d_map_diffusion_coefficient} shows a zoomed-in view of the region surrounding the planet with increased numerical resolution (doubled along each dimension). These effective diffusivity maps trace the regions in which planetary dust stirring is strongest. We find two maxima on opposite sides of the planet at the approximate location where the planetary spiral wake intersects with the edge of the planetary gap, i.e., the location where overshooting horseshoe flows and quasi-circular flows converge, as described in section \ref{sec:Three-dimensional Dust Morphology}. There, the vertically stirred dust is dragged along with the Keplerian flow and carried away from the planet in two opposing directions on almost circular trajectories. The result is the formation of asymmetric features, i.e., two spiral-like arms, originating at the location of the planet and fading as the dust settles downstream. We note that these spiral arm features have a smaller pitch angle than the spiral arms in the gas and tend to become almost circular away from the location of the planet. {For comparison, we trace the spirals in gas with the wake equation of \citet{Rafikov02} (their equation 44) in the first row of \autoref{fig:2d_map_diffusion_coefficient} with white dots (we use their parameters $\nu=0.25$ and $h_p = 0.07$).}\newline
In addition to the azimuthal asymmetry due to the presence of the planet, we also find an asymmetry with respect to the planet itself, with the effective diffusivity being larger in the inner arm than in the outer arm. Apart from the distinct main feature, we find a background distribution that traces spiral features in the outer disk, but with a significantly smaller contrast than the main spiral. \newline
In the third row of \autoref{fig:2d_map_diffusion_coefficient}, we show the corresponding azimuthally averaged effective diffusivity with a solid line and illustrate the one sigma deviations from the average value with the shaded area as a measure of azimuthal variability. In the simulations containing a Jupiter-mass planet with $Sc=100$ (\textit{right}), we find two maxima in the azimuthally averaged effective diffusivity at $\sim0.85r_p$ and $\sim1.25r_p$ with an average value of $\delta_\mathrm{eff}\sim3\cdot10^{-2}$ and $\delta_\mathrm{eff}\sim7\cdot10^{-3}$ respectively. {The azimuthal mean of the inner maximum is almost an order of magnitude above the initial value, while the azimuthal mean of the outer maximum is about a factor of three above the initial value}. Locally, the effective diffusivity is increased by the planet by almost two orders of magnitude, with values peaking above $\delta_\mathrm{eff}>10^{-1}$. As the strength of the background turbulent diffusion increases (decreasing Schmidt number), planetary features in the diffusivity maps become less prominent and are swallowed by the background diffusion. Also, the azimuthal variability decreases. In our simulations with full turbulent diffusion ($Sc=1$, \textit{left}), the diffusivity deviates only marginally from the equilibrium value beyond the immediate planetary region. 

\subsubsection{Influence of the diffusion coefficient}
We find the dust flow morphology to be weakly dependent on the level of background turbulent diffusion (without changing the gas viscosity). The {dust} flow, which is the main driver of the planetary dust stirring, only marginally influences the gas flow via the back reaction due to the dust-to-gas ratio generally being below unity. At the same time, we find the flow structure in the planet's Hill sphere to be highly dynamic and far away from an equilibrium distribution, in agreement with the findings of \cite{Krapp22}. Thus, the timescales responsible for the localized vertical stirring are significantly shorter than the diffusion timescales (also compared to the viscous timescale). At the same time, the vertically extended dust flow that approaches the planet is vertically compressed, along with the gas, as it approaches the Hill sphere due to the increased vertical gravity. Thus, the bulk of the dust approaches the planet on a horseshoe trajectory close to the midplane, regardless of the level of background turbulent diffusion.\newline
As a result of the flow structure being largely independent of the level of background turbulent diffusion, the columnar features visible in the dust density distribution can be drowned in the diffused background distribution if the background disk is thicker than the dust plumes created by the planet. On the other hand, dust structures caused by the planetary stirring become more prominent if the level of background turbulent diffusion is small. Similarly, the plumes in the density distribution become indistinguishable from the background distribution if their vertical extent decreases due to e.g., weaker stirring by a less massive planet. Ultimately, whether the asymmetric features due to planetary stirring stand out in the three-dimensional dust density distribution depends on the relative strength of planetary stirring and background turbulent diffusion. The features are favored in disks with low levels of background turbulent diffusion containing a massive planet.\newline
In our simulations containing a Jupiter-mass planet or below, we find the extent of the vertical dust plumes to be comparable to the planetary Hill radius $r_H$.
{Only in the 5 Jupiter-mass case, we find it to be smaller, which is also the only case in which the Hill radius is significantly larger than the disk scale height ($r_H\simeq2.4h_g$)}.

\subsection{Synthetic Continuum Observations}
\label{sec:Synthetic Continuum Observations}
A natural question that follows from the analysis in the previous sections is, what are the impacts of the discussed dust stirring mechanisms, i.e., planetary stirring and turbulent diffusion on astronomical observations. In this study, we focus on ALMA continuum observations, which trace the thermal emission of the dust component in the disk, with the goal to analyze the effects of turbulent diffusion and planetary dust stirring on continuum observations of protoplanetary disks. To isolate the effect of turbulent diffusion, we first analyze intensity maps of smooth, azimuthally symmetric disks (without an embedded planet) in section \ref{sec:Smooth disks}, before we analyze disks with an embedded Jupiter-mass planet in section \ref{sec:with an embedded planet}. 

\begin{figure*}
\includegraphics[width=1.95\columnwidth]{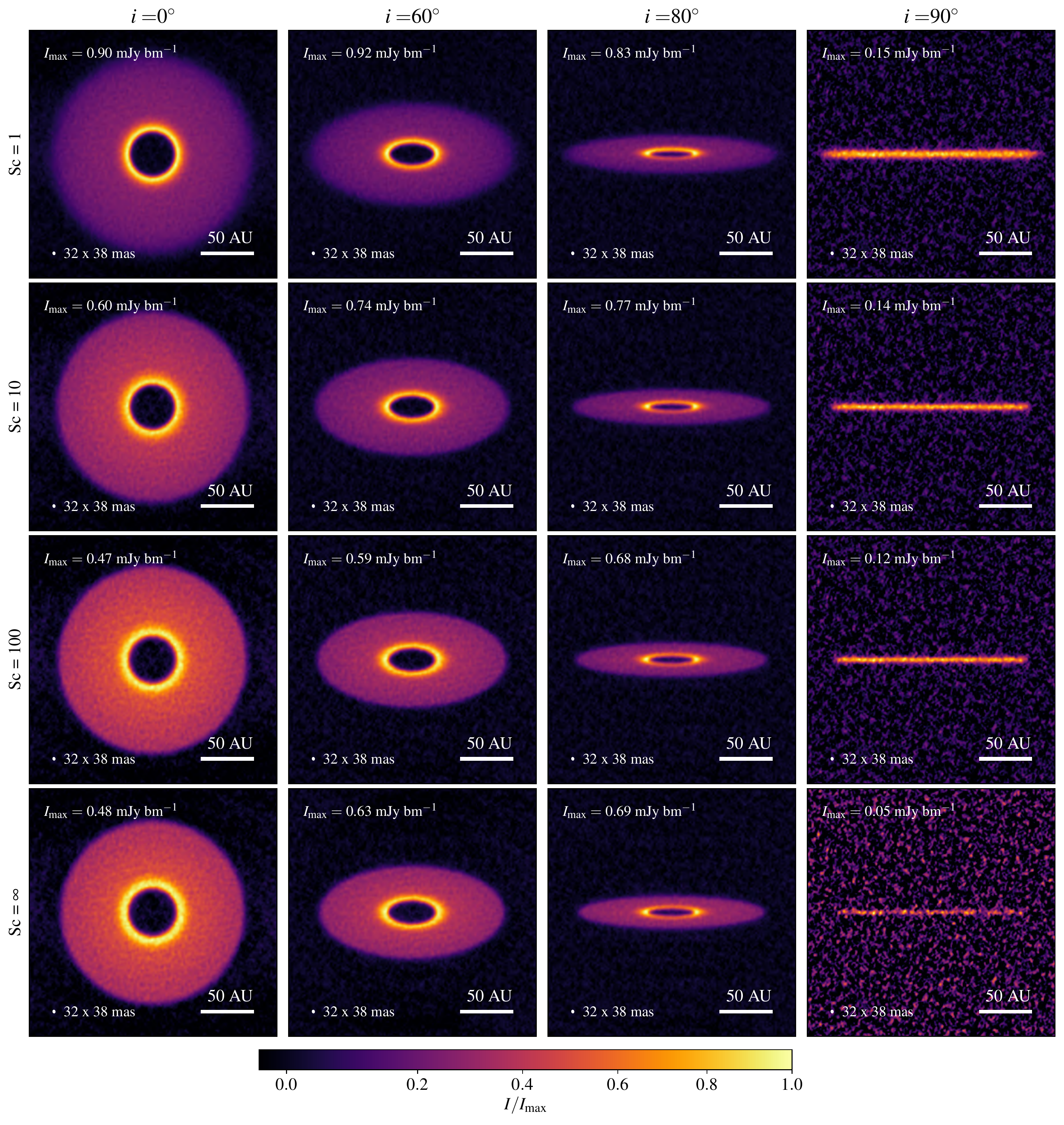}
\caption{Synthetic ALMA Band 7 (345 GHz) observations, obtained with the C43-8 + C43-5 antennae configuration, demonstrating the observational influences of turbulent diffusion and inclination $i$ on axisymmetric disks (\texttt{50au}-domain). Each row shows the same disk at different inclinations $i=0^\circ,60^\circ,80^\circ,90^\circ$. From top to bottom, we show disks with decreasing strength of turbulent diffusion, with the fourth row displaying a simulation without turbulent diffusion. The colormap is normalized to the peak intensity and stretched with a 0.8-power law. The beamsize is indicated in the lower-left corner of each subplot.}
\label{fig:CASA_noplanet_juptemp}
\end{figure*}

\begin{figure*}
\includegraphics[width=1.95\columnwidth]{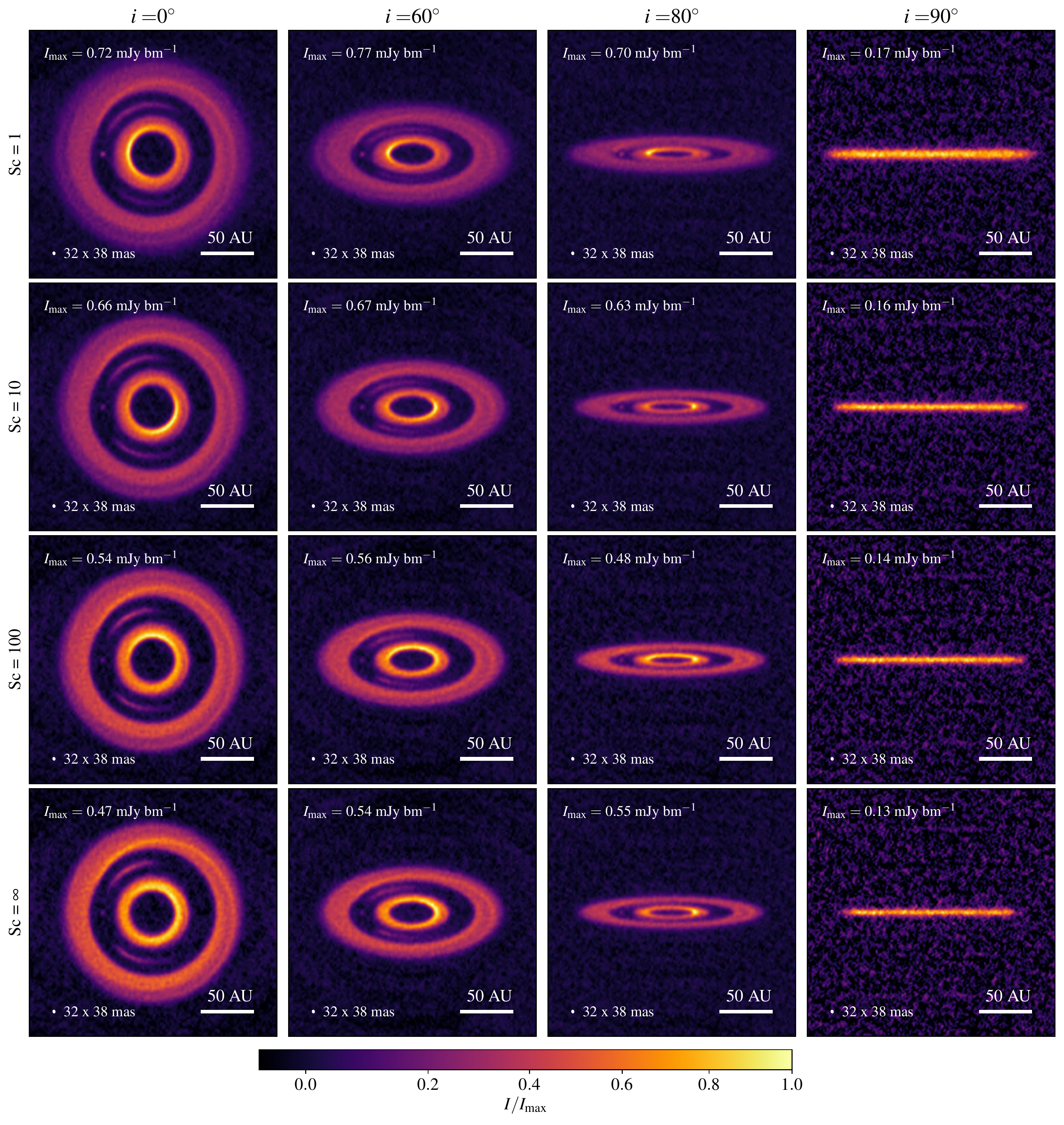}
\caption{Synthetic ALMA Band 7 (345 GHz) observations, obtained with the C43-8 + C43-5 antennae configuration, demonstrating the observational influences of turbulent diffusion and inclination on a disk with an embedded Jupiter-mass planet on a circular orbit at 50 au. Each row shows the same disk at different inclinations $i=0^\circ,60^\circ,80^\circ,90^\circ$. From top to bottom, we show disks with decreasing strength of turbulent diffusion, with the fourth row displaying a simulation without turbulent diffusion. The colormap is normalized to the peak intensity and stretched with a 0.8-power law. The beamsize is indicated in the lower-left corner of each subplot.}
\label{fig:CASA_50AU_Jupiter_mass}
\end{figure*}

\subsubsection{Synthetic observations of axisymmetric disks without planetary perturber}
\label{sec:Smooth disks}
\begin{figure*}
\includegraphics[width=1.95\columnwidth]{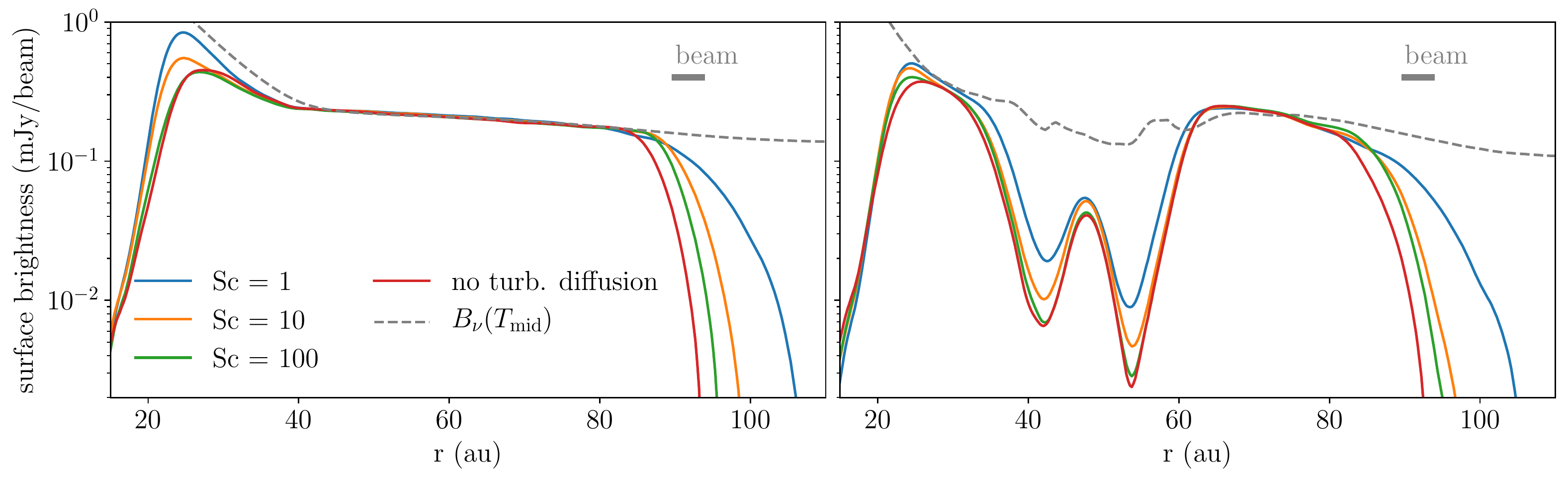}
\caption{Azimuthally averaged radial intensity profile of the face-on view ($i=0^\circ$) of the synthetic ALMA observations of axisymmetric disks (\textit{left}) and disks containing a Jupiter-mass planet at 50 au (\textit{right}) with different strength of turbulent diffusion ($Sc=1,10,100$). The profiles were obtained from the synthetic observations presented in \autoref{fig:CASA_noplanet_juptemp} and \autoref{fig:CASA_50AU_Jupiter_mass}. The gray dashed line represents the Planck function $B_\nu(T)$ evaluated at the midplane temperature $T_\mathrm{mid}$ and $\nu=345$ GHz (band 7).}
\label{fig:noplanet_radial_profile}
\end{figure*}

In \autoref{fig:CASA_noplanet_juptemp} and \autoref{fig:CASA_50AU_Jupiter_mass}, we show synthetic ALMA continuum observations of axisymmetric circumstellar disks (without an embedded planet) with different strength of turbulent diffusion (decreasing from top to bottom) and corresponding disks with an embedded Jupiter-mass planet, respectively. We first focus on \autoref{fig:CASA_noplanet_juptemp}, where, from left to right, the inclination of the disk increases from $i=0^\circ$ in the first column to $i=90^\circ$ in the fourth column. From top to bottom, we decrease the strength of turbulent diffusion and show Schmidt numbers of 1, 10 and 100 in the first, second and third row, respectively. In the fourth row, we show a disk without prescribed turbulent diffusion, in which the dust is pressureless and has completely settled. \newline
%general description - face-on view
In the face-on views of the disks in \autoref{fig:CASA_noplanet_juptemp}, effects of turbulent diffusion are visible at the inner and outer edges of the disks. With increasing strength of turbulent diffusion, the outer edge of the disk diffuses radially outward and counteracts the radial inward drift. This is especially apparent in the face-on view ($i=0^\circ$) and also, but to a lesser degree, in the inclined disks ($i>0^\circ$). In the inner disk, stellar irradiation increases the disk temperature, which in turn increases the diffusion pressure (see equation (\ref{eq:diff_press_def}) and its dependence on equation (\ref{eq:def_pebble_speed}) and the gas sound speed in equation (\ref{eq:stopping_time})). As a result, dust diffuses radially away from the hot inner edge of the disk, and is also more extended vertically. Since the underlying temperature distribution is almost identical in all the presented models in \autoref{fig:CASA_noplanet_juptemp}, the relative difference in the peak intensities between the models, arises solely from the differences in the radial and vertical dust density distribution.\newline
Interestingly, differences in the vertical scale height are not apparent in any view besides the edge-on view ($i>90^\circ$). We will study the edge-on case separately in section \ref{sec:Edge-on disks}. Derived from the intensity maps, we show, in the left sub-plot of \autoref{fig:noplanet_radial_profile}, the azimuthally averaged radial intensity profile of the face-on views ($i=0^\circ$). The differences at the disk edges become apparent. On the other hand, the observed intensity away from the edges of the disk is very similar between the presented models and is closely matching the blackbody emission shown by the Planck function $B_{\nu}(T)$ evaluated at the disk midplane temperature $T_\mathrm{mid}$ and $\nu=345$ GHz (band 7).\newline

\subsubsection{Synthetic observations of disks with an embedded Jupiter-mass planet}
\label{sec:with an embedded planet}
In this section, we study the effects of an embedded Jupiter-mass planet on the observed continuum emission of the same disks, as discussed in the previous section \ref{sec:Smooth disks}.   
%Face-on disks
We present the synthetic continuum observation of our models containing a Jupiter-mass planet on a circular orbit at 50 au (orbiting in a counterclockwise direction) in \autoref{fig:CASA_50AU_Jupiter_mass}.
Like in the previous section, we present four models with varying strengths of turbulent diffusion (Sc = 1,10,100,$\infty$) at four different inclinations ($i=0^\circ,60^\circ,80^\circ,90^\circ$). The images are oriented such that the planet is located at the 9 o'clock position. In this subsection, we focus on the observational disk features caused by the Jupiter-mass planet and aim to link them to the hydrodynamical models. When focusing on the face-on views ($i=0^\circ$), we notice a wide ring outside the orbit of the planet. Like in the no-planet case, the outer edge of the ring radially diffuses outward when turbulent diffusion is active. In contrast, the inner edge of the outer ring, i.e., the edge facing the planet, seems to be only marginally affected by turbulent diffusion and appears equally sharp at different levels of turbulent diffusion. This is supported by the azimuthally averaged radial intensity profile in the sub-plot on the right-hand-side of \autoref{fig:noplanet_radial_profile} where the gap width is slightly narrower only for $Sc=1$. \newline
Moreover, the radial intensity profile just outside the planetary gap is largely flat and does not show the characteristic Gaussian peak which we would expect at the location of the local gas pressure maximum. This indicates that the intensity map traces the dust temperature rather than the dust density distribution, a sign of optically thick emission at the observed wavelength. We confirm this by plotting the Planck function $B_\nu(T)$ to \autoref{fig:noplanet_radial_profile} evaluated at the midplane temperature $T_\mathrm{mid}$ and $\nu=345$ GHz (band 7) which represents fully optically thick emissions and roughly traces the observed intensity profiles. \newline
In all four face-on views, we clearly identify the spiral wake of the planet, peaking in intensity at about 60$^\circ$ behind the planet where the spiral arm emerges from the outer gap edge, i.e., the location where shock heating contributes to a local increase in temperature. With this, we also confirm \cite{Speedie22} and ALMA's potential to detect planetary spirals. \newline
Next, we focus on the emission coming from within the main gap region. We find some emission coming from the location of the planet itself, strongest in the $Sc=1$ case. However, because we do not fully resolve the planetary potential well in our simulations, emission from the planet's vicinity should be followed up with mesh-refined future simulations, thus we will not comment on it further. The second feature we find is emissions from along the co-rotation radius of the planet, with the majority of the emission coming from the location of the two Lagrange points L4 and L5. The population of co-rotating dust is azimuthally more equally distributed in simulations with an increased strength of turbulent diffusion, but the radial extent seems insensitive to the strength of turbulent diffusion. This is mainly because the strength of radial diffusion is reduced in the gap region, where dust grains are only marginally coupled as a result of angular momentum conservation and {epicyclic} oscillations (also see discussion in section \ref{sec:rad_turb_diff}). However, we remark that the observational features within the planetary gap should be taken with a grain of salt because marginally coupled dust grains tend to undergo crossing trajectories, an effect that we currently fail to capture using the fluid approach for dust. However, it is worth noting that the existence of a population of co-rotating dust grains is also predicted by three-dimensional particle-based studies \citep{Fouchet2007,Zhu2014}. Nonetheless, we expect these features to be transient and the planetary co-rotation region away from the two stable Lagrange points to become depleted eventually if we ran the simulations for longer \citep[e.g.][]{Dong2018}. \newline
Similarly to the outer gap edge, the inner gap edge in the intensity maps is not greatly affected by turbulent diffusion. {Unlike in the outer disk, in the synthetic observations in \autoref{fig:CASA_50AU_Jupiter_mass}, we do not identify a prominent planetary spiral in the inner disk.}

\begin{table*}
\begin{center}
\caption{This table shows the optically thin dust masses of the synthetic ALMA observations as calculated with equation (\ref{eq:optically_thin_mass}). The upper half of the table shows values obtained from the axisymmetric disk observations shown in \autoref{fig:CASA_noplanet_juptemp}. The bottom half shows the values obtained from synthetic observations of disks with an embedded Jupiter-mass planet, as shown in \autoref{fig:CASA_50AU_Jupiter_mass}. For models with different strengths of turbulent diffusion, we list, in the third column of the table, the actual dust mass present in the models in units of Earth masses ($M_\mathrm{d,hydro}$). The last four columns list the optically thin dust mass retrieved from the observations at different inclinations in units of Earth masses ($M_{d,i}^\mathrm{thin}$). The measured optically thin disk dust masses are significantly lower than the actual dust masses present in the disk and decrease with increasing inclination. See section \ref{sec:Optically thin dust masses} for a discussion.}
\begin{tabular}{l l c c c c c c} 
 \hline
 \hline
  & model & $M_\mathrm{d,hydro}$ ($M_{\earth}$) & $M_{d,i=0^\circ}^\mathrm{thin}$ ($M_{\earth}$) & $M_{d,i=60^\circ}^\mathrm{thin}$ ($M_{\earth}$) & $M_{d,i=80^\circ}^\mathrm{thin}$ ($M_{\earth}$) & $M_{d,i=90^\circ}^\mathrm{thin}$ ($M_{\earth}$) \\ 
  \hline
    without planet & Sc = 1  & 127 & 73.4 &  36.2 & 13.8 & 2.40\\ 
    & Sc = 10  &  123  & 65.9 &  32.2 & 12.1 & 1.64\\ 
    & Sc = 100 &  122  & 61.7 &  29.9 & 11.2  & 1.18 \\ 
    & Sc = $\infty$ &  119 & 60.1 &  32.9 & 12.0 & 0.34\\ 
     \hline
    with planet & Sc = 1   &  127  & 49.0 &  28.1 & 12.1 & 3.08\\ 
                & Sc = 10  &  123  & 43.0 &  24.6 & 10.1 & 2.18\\ 
                & Sc = 100 &  121  & 42.2 &  24.0 & 9.61  & 1.69 \\ 
           & Sc = $\infty$ &  116  & 38.2 &  22.1 & 8.81 & 1.35\\   
     
 \hline
 \hline
\end{tabular}
\label{table:dustmass_noplanet}
\end{center}
\end{table*}

\begin{figure}
\includegraphics[width=0.95\columnwidth]{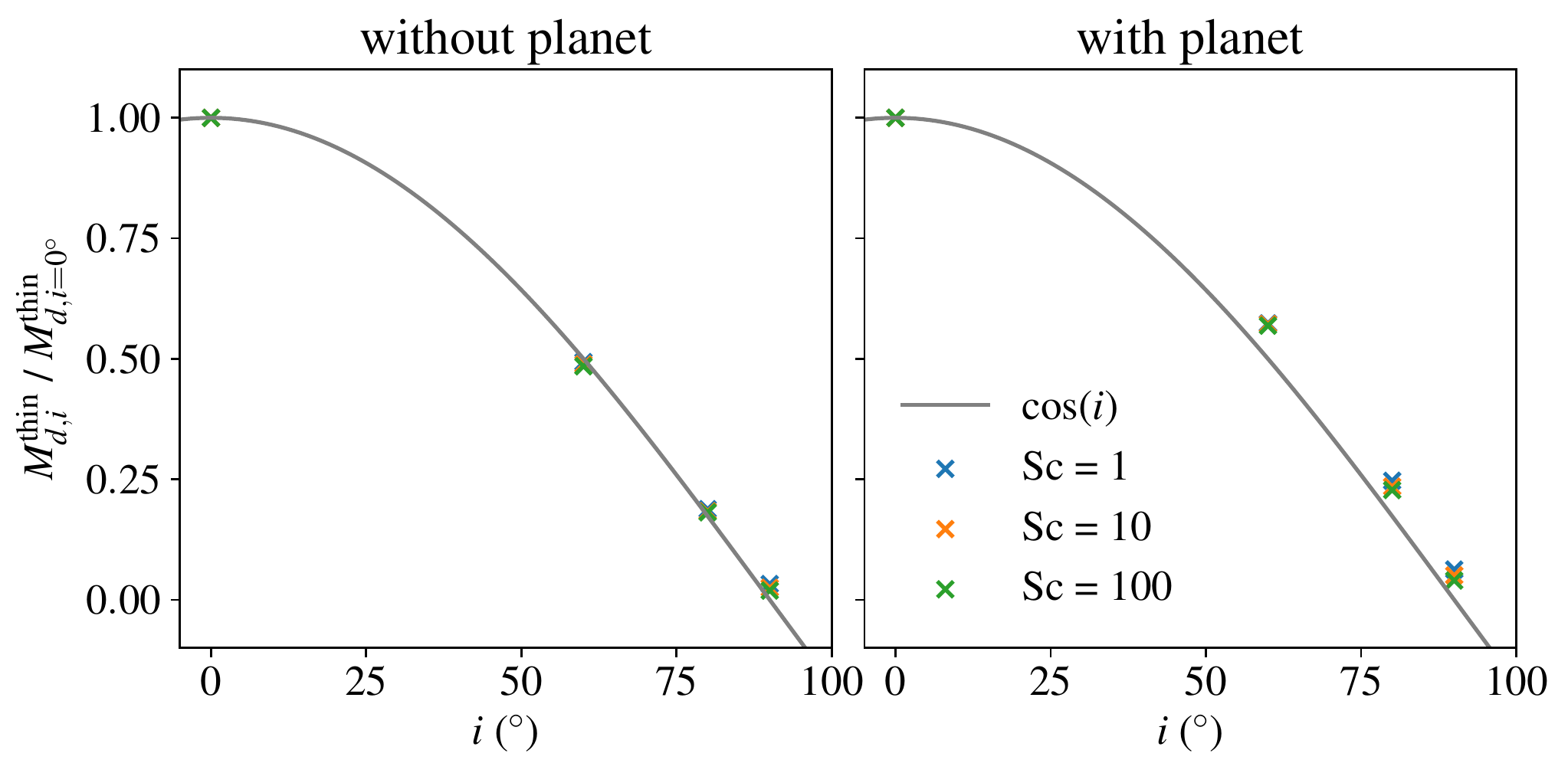}
\caption{Visualization of the dependence of the optically thin dust mass $M_{d,i}^\mathrm{thin}$, obtained with equation (\ref{eq:optically_thin_mass}) and listed in \autoref{table:dustmass_noplanet}, on the inclination $i$. The left-hand side shows the values of axisymmetric disks, and the right-hand side shows the values of the disks containing a Jupiter-mass planet at 50 au. The measured values roughly follow the $\cos(i)$-function, which is represented by the solid gray line in the plots. See section \ref{sec:Optically thin dust masses} for a discussion.}
\label{fig:disk_mass_vs_i}
\end{figure}

\subsubsection{Optically thin dust masses derived from the ALMA mock images}
\label{sec:Optically thin dust masses}
From both sets of observations (without and with an embedded planet), we retrieve the optically thin dust mass $M_d^\mathrm{thin}$ , as routinely done in observational surveys of protoplanetary disks  \citep[e.g.][]{Ansdell2016,Tychoneiec2020}. Here, we aim to study the influence of turbulent diffusion, the presence of a planet and its inclination. From the observed fluxes $F_{\nu=345GHz}$ integrated over the entire disk, we calculate the optically thin dust mass as \citep[][]{Hildebrand1983}
\begin{equation}\label{eq:optically_thin_mass}
    M_{d}^\mathrm{thin}=\frac{F_\nu d^2}{\kappa_\nu B_\nu(T_{d})}
\end{equation}
where $d = 100$ pc is the distance to the source, $\kappa_\nu=10.1\:\mathrm{cm^2/g}$ the absorption opacity (identical to the one used in the radiative transfer calculation), and $B_{\nu}$ the Planck function at the observed frequency $\nu=345$ GHz. We set the dust temperature $T_{d}$ equal to 7.9 K, the mass-averaged midplane temperature in our models.  We summarize our results in \autoref{table:dustmass_noplanet} and list the total dust mass present in the model ($M_\mathrm{d,hydro}$) in the third column in units of Earth mass. In columns four to seven of \autoref{table:dustmass_noplanet}, we list the optically thin dust masses, calculated with equation (\ref{eq:optically_thin_mass}), for the 2$\times$16 synthetic observations presented above. As a result of the marginally optically thick emission in the models, we find the optically thin dust mass $M_d^\mathrm{thin}$ to underestimate the actual dust mass by a factor of 1.7 to 3.0 for the face-on views. Further, we find the optically thin dust mass $M_d^\mathrm{thin}$ to generally decrease with decreasing strength of turbulent diffusion. This is mainly because dust contained in dense optically thick regions diffuses into less dense optically thin disk regions, which increases the observed flux from these regions. We also find the optically thin dust mass $M_d^\mathrm{thin}$ to decrease with increasing inclination $i$. Based on geometrical arguments, one can show that the optically thin dust mass $M_d^\mathrm{thin}$, in an optically thick and geometrically thin disk, decreases with increasing inclination like $\cos(i)$. In \autoref{fig:disk_mass_vs_i}, we plot the normalized optically thin disk dust masses $M_{d,i}^\mathrm{thin}$ as a function of the inclination $i$. The gray line follows the cosine of the inclination ($\cos(i)$). The left subplot of \autoref{fig:disk_mass_vs_i} shows the normalized optically thin disk dust masses of the disks without an embedded planet. The results agree well with the cosine function, indicating that the disk aspect ratio is small, and the disk is mainly optically thick. Only in the edge-on view ($i=90^\circ$), the measured dust masses lie above the cosine function. This is because for very inclined disks, edge effects and the fact that the observed disk is not perfectly geometrically thin become important. Overall, this is an important result, especially for the dust-mass measurement in unresolved, optically thick disks in which the inclination can not be determined. Due to inclination effects, the measured optically thin dust mass in inclined disks is underestimated by a factor $\cos(i)$ with respect to the face-on view. \newline
In the right subplot of \autoref{fig:disk_mass_vs_i}, we plot the optically thin dust masses of the disks containing a Jupiter-mass planet. Like the dust masses in the disk without a planet, the normalized dust masses in the disk with an embedded planet follow the cosine function. However, unlike in the no-planet case, the results for inclined disks are shifted slightly above the cosine line. We expect the difference to arise from optically thin dust in the gap region or and a more vertically extended disk structure caused by the planet stirring. \newline

\begin{figure}
\includegraphics[width=0.95\columnwidth]{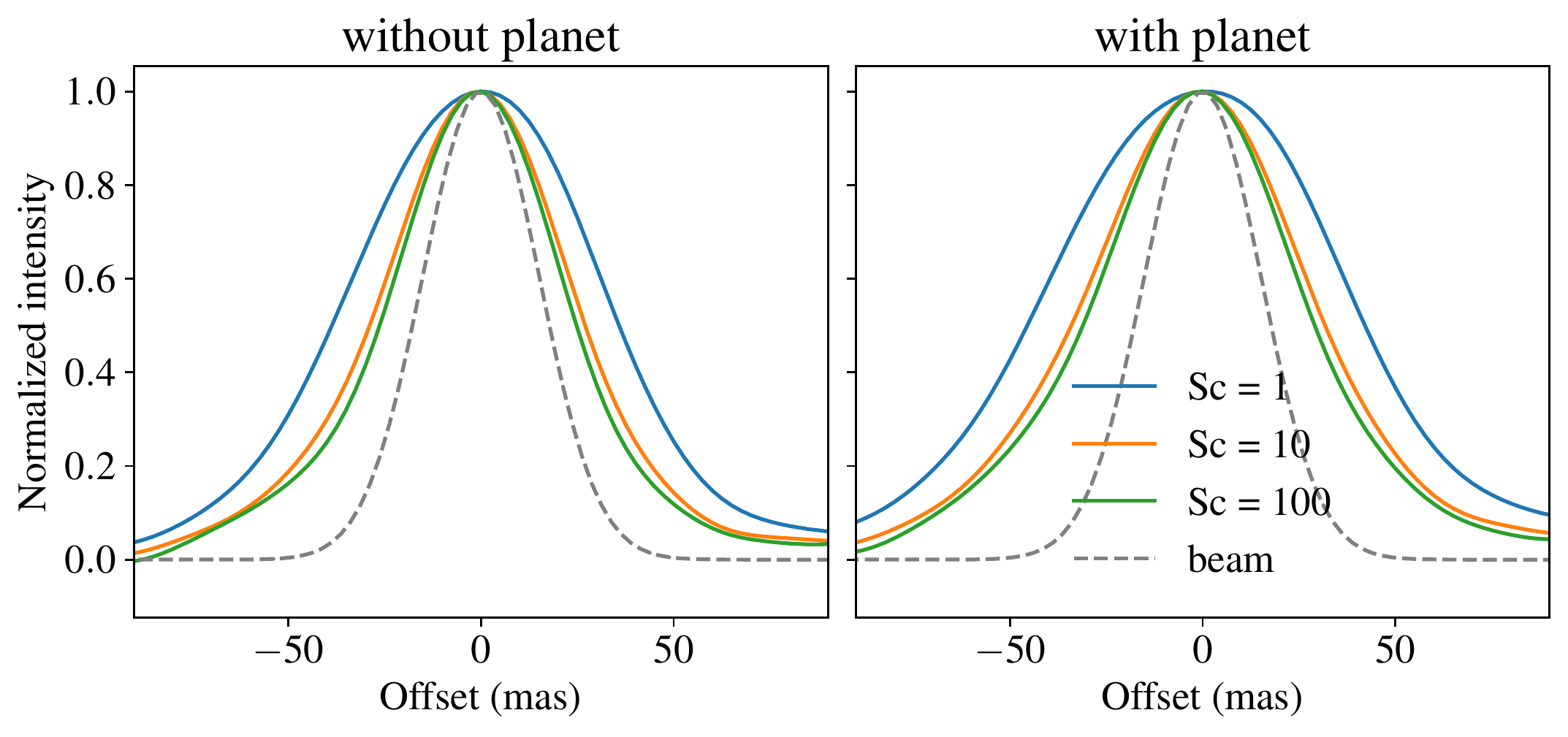}
\caption{Normalized and average minor axis profiles of axisymmetric disks (\textit{left}) and disks containing a Jupiter-mass planet (\textit{right}) obtained from the synthetic observations of the \texttt{50au}-domain in edge-on view ($i=90^\circ$). The different colored profiles represent disks with different strengths of turbulent diffusion. The dashed line indicates a Gaussian profile with FWHM equivalent to the beamsize in the synthetic observations.}
\label{fig:minor_gap_profile}
\end{figure}

\begin{table}
\begin{center}
\caption{Summary of the vertical disk thickness of the axisymmetric disks (\textit{upper half}) and disks with a Jupiter-mass planet (\textit{lower half}) with different strengths of turbulent diffusion (decreasing from top to bottom). The third column lists the measured deconvolved disk scale heights ($w_m$) of the underlying disk profile as measured in the edge-on view $(i=90^\circ)$ of the synthetic observations and  calculated with equation (\ref{eq:deconv}). The fourth column lists the corresponding scale heights ($w_\mathrm{rt}$) measured directly after the radiative transfer calculation, i.e., before the convolution with the ALMA beam and the addition of the thermal noise. The second to last column lists the azimuthally averaged dust scale height ($h_{d,\mathrm{80\:au}}$) measured in the underlying hydrodynamic simulation at a heliocentric distance of 80 au. The last column lists the ratio of the true underlying scale height with the hydrodynamic scale height  ($w_\mathrm{rt}/ h_{d,\mathrm{80\:au}}$). The former values are generally larger by a factor of a few.}
\begin{tabular}{l l c c c c c} 
 \hline
 \hline
  & Sc & $w_m$ (au) & $w_\mathrm{rt}$ (au) & $h_{d,\mathrm{80\:au}}$ (au) & $\frac{w_\mathrm{rt}}{ h_{d,\mathrm{80\:au}}}$\\ 
  \hline
    without & 1   & 2.96 &  2.99 & 1.20 & 2.49\\ 
    planet & 10 & 1.81 &  1.97 & 0.73 & 2.70\\ 
    & 100 & 1.54 &  1.40 & 0.52  & 2.70 \\ 
    & $\infty$ & 1.18 &  0.43 & 0.44 & 0.97\\ 
    \hline
    with & 1   & 3.83 &  3.23 & 1.34 & 2.69\\ 
                 planet & 10 & 2.27 &  2.17 & 0.80 & 2.97\\ 
                & 100 & 2.00 &  1.55 & 0.55  & 2.98 \\ 
           & $\infty$ & 1.92 &  1.19 & 0.45 & 2.70\\

 \hline
 \hline
\end{tabular}
\label{table:map_measurements}
\end{center}
\end{table}

\subsubsection{Synthetic observations of edge-on circumstellar disks}
\label{sec:Edge-on disks}
In the previous sections, we have discussed how turbulent diffusion affects the inner and outer edges of the disks, where sharp edges are radially smeared out which has a direct impact on optically thin dust masses. However, the effects of turbulent diffusion on the vertical extent of the disk are hardly discernible in the disks with low inclinations ($i<90^\circ$). We now focus on the edge-on observations ($i=90^\circ$) presented in \autoref{fig:CASA_noplanet_juptemp} and \autoref{fig:CASA_50AU_Jupiter_mass} in which we aim to analyze effects of turbulent diffusion and planetary stirring in the vertical direction. Because the effective diffusivity $\delta_\mathrm{eff}$ is closely related to the vertical dust scale height $h_d$, we aim to infer the vertical extent of the disk from the edge-on observations and relate it to the strength of vertical stirring (similar to the procedures presented in \cite{Villenave20,Vilenave22}). For this, we first estimate the resolution required to resolve the dust scale height $h_d$ along the minor axis of a disk in an edge-on view. Assuming the dust scale height is approximately Gaussian with scale height $h_d$, the FWHM, i.e., the beam size of the observing beam, must be smaller than $\sim2.355 h_d$ (to fully resolve the scale height, the beam should sample this value at least twice). In the observations presented in  \autoref{fig:CASA_noplanet_juptemp} and \autoref{fig:CASA_50AU_Jupiter_mass}, using the ALMA configuration C43-8 concatenated with C43-5, the mean beam size is 35 mas. Therefore, based on this rough estimate, at a distance of 100 pc, dust scale heights above $h_d\gtrsim 1.5$ au can be properly resolved. Assuming, the optical surface that we observe in the edge-on view lies at about 80 au from the central star, this corresponds to a disk aspect ratio of $h_d/r = 0.019$. In our models, the gas disk has an aspect ratio $h_g/r\sim 0.05$ at 80 au. We then use equation (\ref{scale_height_ratio}) to obtain an approximate lower limit of the ratio between diffusivity $\delta$ and the Stokes number $St$ that are required for us to resolve the disk vertically. We find $\delta/St>0.14$. At 80 au, the midplane Stokes number in our models is $St_\mathrm{mid} \sim 0.065$. Hence, we expect the disk to be resolved along its minor axis in the edge-on view if $\delta > 9.1\cdot 10^{-3}$. Therefore, we expect the disk scale height in the models with $Sc\leq10$ to be properly resolved in ALMA observations. This might be improved when a new generation of radio interferometers with even better angular resolution {becomes available}. For example, ngVLA is expected to resolve sub-mas-scales \citep[][]{Selina18}, {i.e., an order of magnitude better compared to the capabilities of ALMA. Thus, as long as the upper end of the dust size distribution remains comparable or smaller than ALMA wavelengths, we expect to observe more optically thin emissions at a higher angular resolution with ngVLA compared to ALMA. However, if the upper end of the dust size distribution extends into the regime comparable to ngVLA wavelengths $\lambda_\mathrm{obs.}$ ($6-300$ mm compared to $0.3-3$ mm for ALMA), the advantage might only be marginal at best. This is because, assuming the observed particle size is proportional to the observed wavelength, the scale height of the observed particles scales inversely with the observed wavelength $h_d\propto \lambda_\mathrm{obs.}^{-1}$ and thus, the larger particles probed at ngVLA wavelengths are more settled compared to the particles probed with ALMA.} \newline
In \autoref{fig:minor_gap_profile}, we present the average brightness profiles (averaged along the central 0.5 as) along the minor axis of the synthetic edge-on disk observations without a planet (left) and with an embedded Jupiter-mass planet (right). The dashed line corresponds to a Gaussian beam profile with FWHM = 35 mas. We measure the scale height of the observed minor axis $\sigma_m$ by fitting a Gaussian to the profiles. Because the profiles are convolved with the ALMA beam, we have to deconvolve to obtain the true scale height of the underlying profile $w_m$. Assuming the underlying profile is also Gaussian, the true scale height of the underlying minor axis profile can be found using the formula
\begin{equation}\label{eq:deconv}
    w_m = \sqrt{\sigma_m^2-\sigma_b^2}
\end{equation}
where $\sigma_b$ is the standard deviation of the beam. We list the values of the deconvolved scale height $w_m$ in \autoref{table:map_measurements}. Comparing them to the beamsize (1.5 au), only the minor axis of the model in which turbulent diffusion is strongest ($Sc=1$) and contains an embedded planet is sampled by the beam at least twice. All the other models are vertically not well resolved. But, only the model without prescribed turbulent diffusion ($Sc=\infty$) vertically extends less than one beam. Nevertheless, we evaluate the influence of the convolution/deconvolution and also measure the average FWHM of the minor axis profile in the output of the radiative transfer calculation ($\mathrm{FWHM}_\mathrm{rt}$) to estimate the actual underlying scale height with $w_\mathrm{rt} = \mathrm{FWHM}_\mathrm{rt} / 2.355$. We list the values of the estimated underlying scale heights $w_\mathrm{rt}$ in the third column of \autoref{table:map_measurements}. We find the measured scale heights $w_m$ to agree well with the true underlying scale heights $w_\mathrm{rt}$ if they are comparable or larger than the beam size (1.5 au). However, the measured scale heights $w_{rt}$ are larger by a factor of 2.49-2.98 compared to the hydrodynamical scale heights evaluated at 80 au $h_{d,80\:au}$ (obtained from the FWHM). We list the detailed value for each model in the sixth column of  \autoref{table:map_measurements}. The difference between the two values is a result of the large optical depths along the line of sight in the edge-on view. 

%%%%%%%%%%%%%%%%%%%%%%%%%%%%%%%%%%%%%%%%%%%%%%%%%%%%%%%%%%%%%%%%%%%%%%%%%%%%%%%%%%%%%%%%

%%%%%%%%%%%%%%%%%%%%%%%%%%%%%%%%%%%%%%%%%%%%%%%%%%%%%%%%%%%%%%%%%%%%%%%%%%%%%%%%%%%%%%%%

\section{Discussion, Summary, and Conclusion}
\label{sec:Discussion}
\subsection{Caveats}
In this work, we incorporate a large amount of the relevant physics important for studying the underlying problem. More specifically, we study the problem in three dimensions and model the main thermodynamic processes that account for the heating and cooling of the protoplanetary disk (radiative, viscous, adiabatic). Further, we model the dynamics of the gas and solid components independently and let them interact via aerodynamic drag. In the solids, we include a {linear and angular} momentum-conserving subgrid model for the turbulent stirring of dust grains. The subsequent radiative transfer step accounts for frequency-dependent interaction between the protoplanetary disk material and radiation. Nonetheless, our approach can benefit from a few improvements, of which some were already listed in section 4.2 of \cite{Binkert21}. Among the most relevant is the fact that we use only a single dust grain size and do not incorporate other grain sizes or the effects of grain growth. We aim to overcome this drawback in a subsequent study. \newline
{We also highlight that the low mass-averaged midplane temperature of 7.9 K, as stated in section \ref{sec:Optically thin dust masses}, is a result of efficient cooling via the disk surface to the 2.7 K background (i.e., the cosmic microwave background). The background radiation in a molecular cloud is likely closer to $\sim10$~K \citep[e.g.][]{Schnee09}. Thus, disks embedded in a molecular cloud core are likely warmer and thus thicker and more strongly flared than the isolated disks considered in this work. \newline}
Further, we keep the planet in our simulations on a fixed circular orbit. In reality, planets are likely to undergo migration, an effect that we do not capture in this work. As outlined in \cite{Binkert21}, especially the rapid type III migration could be of importance \citep[][]{Masset2003} and potentially affect the disk morphology \citep[e.g.][]{Weber20}. Moreover, the presence of multiple planets could also affect the disk morphology.

\subsection{Summary}
Using three-dimensional radiative two-fluid (gas + 1 mm size dust) hydrodynamic simulations of circumstellar disks with an embedded planet, we investigate planet-induced dust stirring in a turbulent background. We model the effects of turbulent diffusion as a pressure-like term in the momentum equation of the dust fluid, which consequently conserves {linear and angular} momentum in our model and  implicitly captures the effects of orbital oscillations, also within the planetary Hill sphere (unlike the conventional mass diffusion model). We further study observational signatures of planet-induced dust stirring and turbulent diffusion in synthetic ALMA (sub-)mm-continuum observations. Our main findings are the following: 

\begin{itemize}
    \item We model turbulent diffusion as a pressure-like term in the otherwise pressureless dust fluid. This approach has the advantage that, as opposed to adding the diffusion flux only to the mass conservation equation, fully conserves {linear and angular} momentum, a vital property to study the protoplanetary disk problem. As a result of the {angular} momentum conservation, the turbulent diffusion model implicitly captures the effects of in-plane epicyclic oscillations as predicted by \cite{Youdin2007}, i.e., the weakening of dust turbulent diffusion in moderately coupled environments ($St\gtrsim 1$). Due to the implicit nature, this is also the case if the flow deviates from being purely Keplerian, e.g., inside the Hill sphere of an embedded planet. As a result, we also accurately capture the diffusive redistribution of dust mass and angular momentum (i.e., the dust accretion behavior) inside the planetary Hill sphere and a potential circumplanetary disk (CPD), in moderately coupled environments of a gap-opening planet. A property that the pure mass diffusion approach does not have. Our model also correctly reproduces the dust distribution in the vertical settling-diffusion equilibrium in both the strong-coupling limit ($St\ll 1$) and the weak coupling limit ($St\gg 1$). In practice, modeling the dust component as a fluid with non-zero pressure also simplifies the numerical treatment because specialized pressureless solvers can be omitted and standard (gas) fluid solvers can be used instead. \newline

    \item In our three-dimensional radiative hydrodynamic simulations, we identify distinct flow structures in the millimeter-size dust in the surroundings of a giant planet, which are driven by the meridional circulation of the gas. {These dust flow structures are inherently three-dimensional, vary strongly in space, leading to disk asymmetries.} We find these distinct flow structures to be only marginally affected by background turbulent diffusion. \newline
    
    \item We quantify the planetary dust stirring by measuring an effective diffusivity $\delta_\mathrm{eff}$ and find it to vary strongly both radially and azimuthally, with two distinct maxima at two opposing sides of the planet. We find the planetary mixing to produce azimuthally averaged values in the range $\delta_\mathrm{eff}\sim 5\cdot 10^{-3}-2\cdot10^{-2}$ and local peaks with values up to $\delta_\mathrm{eff}\sim 3\cdot 10^{-1}$ for a Jupiter-mass planet. The effective diffusivity scales with the planet mass and if the diffusivity of the background turbulent is large enough in relation to the planet mass, i.e., the turbulent alpha is larger than the effective diffusivity ($\alpha>\delta_\mathrm{eff}$), the effect of planetary stirring on the vertical dust scale height is \textit{drowned} in the turbulently diffusive background. \newline
    
    \item In our synthetic ALMA continuum observations, we find the angular resolution of ALMA to only be sufficient to resolve the vertical structure of our model if turbulent diffusion is relatively strong. We were unable to make the planetary-induced vertical dust structures visible in the synthetic observations because we found the requirements on angular resolution and contrast/sensitivity to be too stringent for ALMA. {In near face-on observations, we confirm \cite{Speedie22} and the capability of ALMA to detect planetary spirals, at least the outer arm.} \newline 
    
    \item We find the total disk mass observationally obtained from using the optically thin approximation to be only weakly affected by the strength of turbulent diffusion. However, as a result of the disks being marginally optically thick, the optically thin dust masses decrease with inclination $i$ and is roughly proportional to $\cos(i)$.\newline
    
    \item In an edge-on observation ($i=90^\circ$), the observed vertical disk scale height of the millimeter continuum emission of the disk overestimates the underlying hydrodynamic scale height by a factor of $\sim 2.5-3$. Not accounting for this difference could result in an overestimation of the vertical strength of turbulent diffusion in real observations. 
\end{itemize}

\section*{Acknowledgements}
We thank the anonymous referee for their valuable comments and suggestions. These results are part of a project that has received funding from the European Research Council (ERC) under the European Union’s Horizon 2020 research and innovation programme (Grant agreement No. 948467). Computations partially have been done on the "Piz Daint" machine hosted at the Swiss National Computational Centre. T.B. acknowledges funding from the European Research Council (ERC) under the European Union's Horizon 2020 research and innovation programme under grant agreement No 714769. F.B. and T.B. acknowledge funding from the Deutsche Forschungsgemeinschaft under Ref. no. FOR 2634/1 and under Germany's Excellence Strategy (EXC-2094–390783311).

\section*{Data availability}
The data underlying this article will be shared on reasonable request to the corresponding author.

%%%%%%%%%%%%%%%%%%%%%%%%%%%%%%%%%%%%%%%%%%%%%%%%%%



\bibliographystyle{mnras}
\bibliography{references} % if your bibtex file is called example.bib

%\begin{thebibliography}{99}
%\end{thebibliography}

%%%%%%%%%%%%%%%%%%%%%%%%%%%%%%%%%%%%%%%%%%%%%%%%%%

%%%%%%%%%%%%%%%%% APPENDICES %%%%%%%%%%%%%%%%%%%%%
\appendix
\label{sec:appendix}

\section{Turbulent Diffusion: Linear Perturbation Analysis}
In this section, we study the properties of the dust diffusion model introduced in section \ref{sec:diffusion_pressure_model} in which we model turbulent diffusion as a pressure in the dust fluid and compare it to the mass diffusion model which models turbulent diffusion with a diffusion term in the mass conservation equation. We do this by performing a linear perturbation analysis in one-dimensional Cartesian coordinates ($x$) in section \ref{sec:One Dimension in the Absence of External Forces} and in a two-dimensional Keplerian disk setup in polar coordinates ($r,\phi$) in section \ref{sec:Two-Dimensional Diffusion in a Keplerian Disk}. At the same time, we use the derived analytical solutions to test our numerical implementation. The first system of equations that we study are equations (\ref{eq:dust_cont_eq2}) and (\ref{eq:wo_external_forces}), and the second system is described by equation (\ref{eq:adv_diff_equation}) and equation (\ref{eq:classical_velocity_equation3}).

\begin{figure*}
\includegraphics[width=1.9\columnwidth]{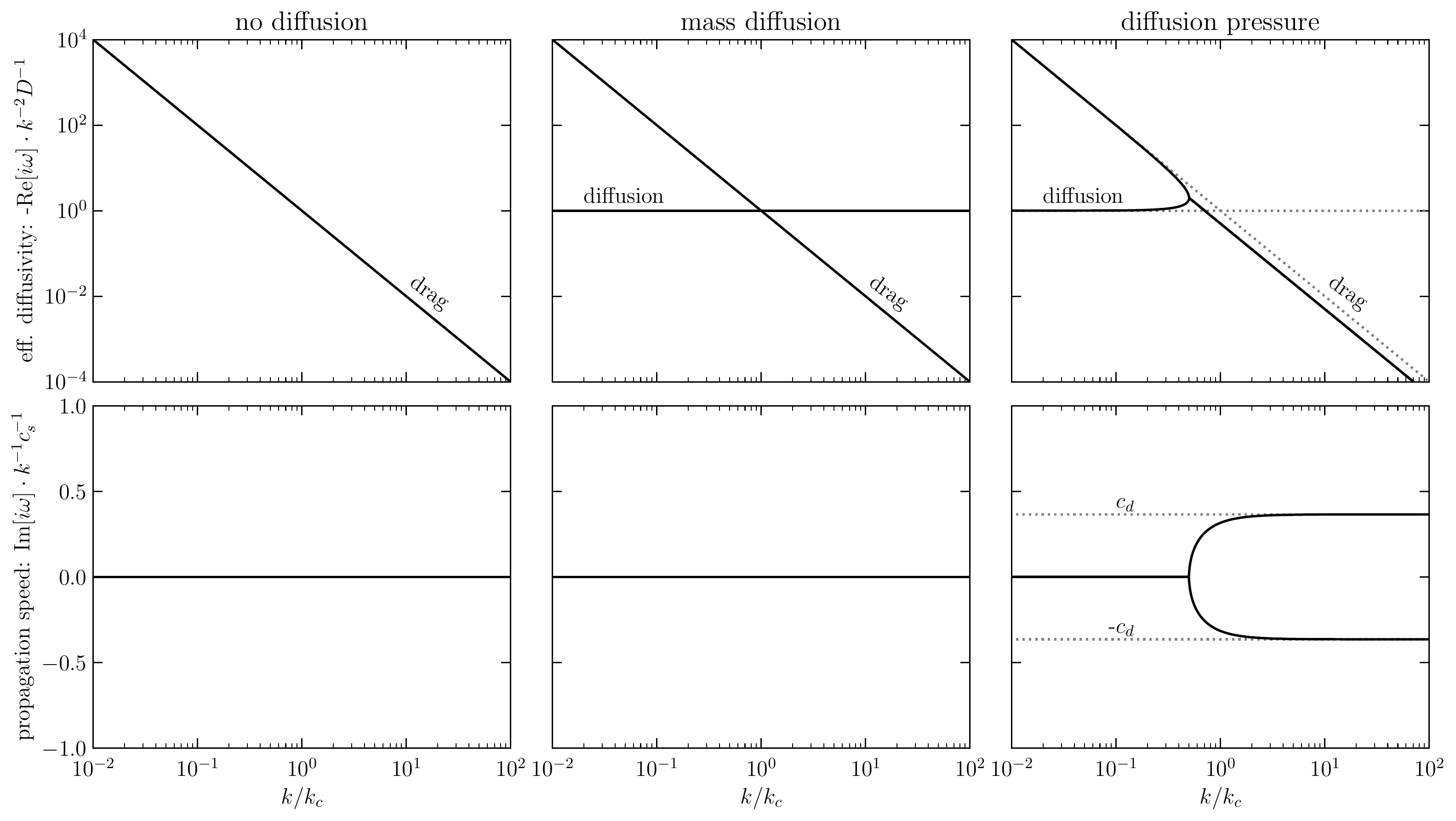}
\caption{Dispersion relations of the one-dimensional linear perturbation analysis in the three turbulent diffusion models, without diffusion (\textit{left}), mass diffusion (\textit{middle}), diffusion pressure (\textit{right}). The upper row displays the real part of the perturbation frequency $\mathrm{Re}[i\omega]$. The data is multiplied by $k^{-2}$ and normalized by $D^{-1}$. Thus, any diffusive solution is a horizontal line and the drag solution, which is independent of the length scale $k$, has a slope $-2$. The lower row displays the imaginary part of the perturbation frequency $\mathrm{Im}[i\omega]$. The data is multiplied by $k^{-1}$ and normalized by $c_s^{-1}$. Thus, any solution that propagates with constant speed is represented by a horizontal line. The model without diffusion (\textit{left}) does not have a diffusion solution. The two models which include diffusion are identical on large scales ($k\gg k_c$). On small scales ($k\ll k_c$), the diffusion solution in the diffusion pressure model merges with the drag solution. There, the perturbation propagates with speeds $\pm c_d$. An effect that is not present in the mass diffusion model.}
\label{fig:disprelation_nodiff}
\end{figure*}

\subsection{One Dimension in the Absence of External Forces}
\label{sec:One Dimension in the Absence of External Forces}
We first perform a one-dimensional linear perturbation analysis on the two systems of equations in the absence of external forces ($\nabla\Phi=0$) and assume the background fields to be static and constant in space. Further, we assume the dust diffusion coefficient $D$ to be constant, and the dust-to-gas ratio to be small $\rho_d/\rho_g\ll 1$ and thus neglect the dust feedback onto the gas. Further, we set $c_s\gg c_d$ such that $c_d^2=D/\tau_s$. On top of the stationary background fields, we introduce small harmonic perturbations on the dust density $\rho_d$ and velocity $v_d$ of frequency $\omega$ and wavelength $\lambda = 2\pi /k$, where $k$ is the wavenumber of the perturbation. Thus, the dust density and velocity have the form $\rho_d(x,t)=\rho_0+\delta\rho(x,t)$ and $v_d(x,t)=\delta v(x,t)$ with $\delta\rho(x,t)=\delta\rho_0\exp({i\omega t + ikx})$ and $\delta v(x,t)=\delta v_0\exp({i\omega t + ikx})$. We assume the background velocity to be zero $v_0=0$ and the perturbation amplitudes to be small $\delta \rho_0 \ll \rho_0$. \newline

\begin{figure*}
\includegraphics[width=1.9\columnwidth]{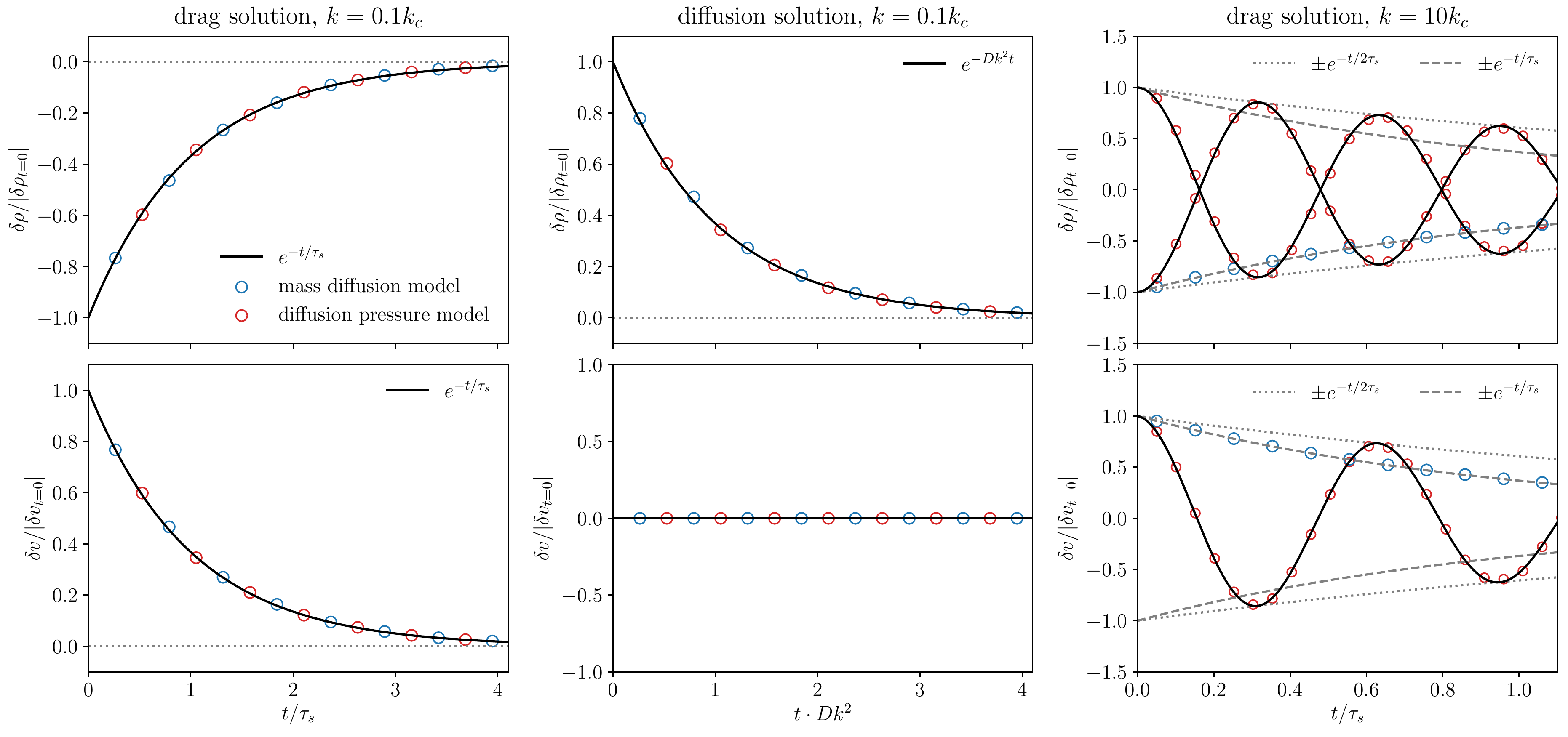}
\caption{ Density (\textit{upper row}) and velocity (\textit{lower row}) eigenstates versus time of the one-dimensional perturbation analysis. Plotted are analytical results (back lines) and numerical results of the mass diffusion model (blue circles) and the diffusion pressure model (red circles) at $x=\pi/4k$ (\textit{first, second column}) and $x=0$ (\textit{third column}) with $\tau_s=\rho_0=1$, $D = 4$ and $\delta \rho_0=\delta v_0=10^{-3}$ (in code units). The left and middle sub-plots show the exponential decay of the drag solution, and the diffusion solution respectively, on a large spatial scale ($k = 0.1k_c$). Both diffusion models agree on this scale. The right sub-plots visualize the difference between the two diffusion models on a small spatial scale ($k = 10 k_c$). While the mass diffusion model is static and decays monotonically and exponentially on a timescale $\tau_s$, the diffusion pressure model propagates wavelike with speed $c_d$ and decays on a timescale $2\tau_s$. Note, for readability, the diffusion solution of the mass diffusion model is not plotted in the right sub-panels. On this small spatial scale, it decays on a timescale $400$ times shorter than the drag solution.}
\label{fig:perturbed_solutions}
\end{figure*}

\subsubsection{1-D Without Diffusion ($D=0$)}
\label{sec:1-D without diffusion}
We first study the systems in the absence of turbulent diffusion ($D=0$) and set the stopping time equal to unity $\tau_s=1$. It is straightforward to see that, in the absence of turbulent diffusion, system (\ref{eq:dust_cont_eq2})/(\ref{eq:wo_external_forces}) is identical to the system (\ref{eq:adv_diff_equation})/(\ref{eq:classical_velocity_equation3}) and its linearized form can be represented by a two-dimensional matrix equation
\begin{equation}
-i\omega
    \begin{pmatrix}
 \delta \rho_d\\
\delta v_d
\end{pmatrix}
= 
    \begin{pmatrix}
0 & ik\rho_0 \\
0 & \tau_s^{-1}
\end{pmatrix}
    \begin{pmatrix}
 \delta \rho_d\\
\delta v_d
\end{pmatrix}
\end{equation}
The dispersion relation of the above equation is
\begin{equation} 
    i\omega(\tau_s^{-1}+i\omega)=0
\end{equation}
which has two solutions of which one is proportional to the inverse of the stopping time
\begin{equation}
\label{eq:drag_solution_no_diff}
    i\omega_1=-\tau_s^{-1}
\end{equation}
This solution represents a perturbation that decays on a timescale equal to the stopping time due to damping by aerodynamic drag. Because of this behavior, we call this solution the \textit{drag solution} of the linearized system. The second solution of the system without diffusion is trivially zero:
\begin{equation}
    i\omega_2=0
\end{equation}
This solution represents a perturbed state which is static, i.e., the velocity perturbation vanishes $\delta v_0=0$. In the first column of \autoref{fig:disprelation_nodiff}, we plot the real (\textit{upper row}) and imaginary part (\textit{lower row}) of the drag solution as a function of the wave number $k$. Note that the upper row of \autoref{fig:disprelation_nodiff} is multiplied by $k^{-2}$ and normalized by the diffusion coefficient $D$. Thus, any diffusive solution is represented by a horizontal line in these plots. The drag solution is distinctly different from a diffusive solution and is independent of the wave number $k$. It has a slope of $-2$ in these plots. The solutions in the lower rows of \autoref{fig:disprelation_nodiff} are multiplied by $k^{-1}$ and normalized by the gas sound speed $c_s$. Thus, any solution which propagates with constant speed is represented by a horizontal line in these plots. The two perturbed eigenstates, corresponding to velocity and density of the drag solution (\ref{eq:drag_solution_no_diff}) are:
\begin{equation}
    \label{eq:vel_wo_diff_ana_solution}
   \mathrm{Re}(\delta v_1) = \delta v_0\cos(kx)e^{-t/\tau_s}
\end{equation}
\begin{equation}
\label{eq:dens_wo_diff_ana_solution}
    \mathrm{Re}(\delta \rho_1) =-\delta v_0 k\rho_0\tau_s\sin(kx)e^{-t/\tau_s}
\end{equation}

\subsubsection{1-D Mass Diffusion}
The second system we analyze is the system in which diffusion is included as a source term in the mass equation. I.e. the dynamics of the dust density is governed by the system (\ref{eq:adv_diff_equation})/(\ref{eq:classical_velocity_equation3}). The velocity equation remains unchanged compared to the system without diffusion. We set $D=1$ and perform the same linear perturbation analysis as in section \ref{sec:1-D without diffusion} and write the linearized system in its matrix representation
\begin{equation}
\label{eq:system_1D-massdiff}
-i\omega
    \begin{pmatrix}
 \delta \rho_d\\
\delta v_d
\end{pmatrix}
= 
    \begin{pmatrix}
Dk^2 & ik\rho_d \\
0 & \tau_s^{-1}
\end{pmatrix}
    \begin{pmatrix}
 \delta \rho_d\\
\delta v_d
\end{pmatrix}
\end{equation}
This system has the following dispersion relation: 
\begin{equation}
    (Dk^2+i\omega)(\tau_s^{-1}+i\omega)=0
\end{equation}
The first solution to this dispersion relation is identical to the drag solution in the previous system without diffusion: 
    \begin{equation}
    \label{eq:drag_solution_mass_diff}
        i\omega_1=-\tau_s^{-1}
    \end{equation}
The second solution is
    \begin{equation}
    \label{eq:diff_solution_mass_diff}
        i\omega_2=-Dk^2
    \end{equation}
which is proportional to $k^2$. Because it has the properties of a diffusion process, we call this solution the \textit{diffusion solution}. We illustrate the two solutions of this system in the second column of \autoref{fig:disprelation_nodiff}. On large spatial scales (small $k$), above the characteristic length scale $\lambda_c=2 \pi/k_c=2 \pi \sqrt{D\tau_s}$, the drag solution decays faster than the diffusion solution. On spatial scales smaller than $\lambda_c$ (large $k$), perturbations decay faster by diffusion than by drag. The set of eigenstates of the linearized system in equation (\ref{eq:system_1D-massdiff}) which belongs to the drag solution (\ref{eq:drag_solution_mass_diff}) are: 
\begin{equation}
\label{eq:vel_mass_diff_ana_solution_2}
   \mathrm{Re}(\delta v_1) = \delta v_0\cos(kx)e^{-t/\tau_s}
\end{equation}
\begin{equation}
\label{eq:dens_mass_diff_ana_solution_2}
    \mathrm{Re}(\delta \rho_1) =\delta v_0\frac{ k\rho_0}{Dk^2-\tau_s^{-1}}\sin(kx)e^{-t/\tau_s}
\end{equation}
These solutions are plotted as a function of time in the first column of \autoref{fig:perturbed_solutions} for $k = 0.1  k_c$ at $x=\pi/4k$. The upper sub-plot shows the normalized density perturbation and the lower sub-plot shows the normalized velocity perturbation. The black solid lines represent the corresponding analytic solution from equation (\ref{eq:vel_mass_diff_ana_solution_2}) and (\ref{eq:dens_mass_diff_ana_solution_2}) respectively, which exponentially decay on a timescale equal to the stopping time $\tau_s$. The blue circles represent the numerical solution for $\tau_s=\rho_0=1$, $D = 4$ and $\delta \rho_0=\delta v_0=10^{-3}$ (in code units) obtained with the \textsc{Jupiter} code. This solution is also plotted in the third column of \autoref{fig:perturbed_solutions} for $k=10 k_c$ (black dashed line). 
\newline
Equivalently to the drag solution, the set of eigenstates that belongs to the diffusion solution (\ref{eq:diff_solution_mass_diff}) are: 
\begin{equation}
   \mathrm{Re}(\delta v_2) = 0
\end{equation}
\begin{equation}
    \mathrm{Re}(\delta \rho_2) =\delta \rho_0\cos(kx)e^{-Dk^2t}
\end{equation}
These solutions are plotted as a function of time in the second sub-panels of \autoref{fig:perturbed_solutions}, also for $k = 0.1  k_c$ at $x=\pi/4k$. The density perturbation of this static solution ($\delta_v = 0$) decays on a timescale equal to $1/Dk^2$.

\subsubsection{1-D Diffusion Pressure}\label{sec:1-D diffusion pressure}
The third system we analyze is the linearized system (\ref{eq:dust_cont_eq2})/(\ref{eq:wo_external_forces}), in which diffusion is modeled by a diffusive pressure term in the velocity equation. The matrix representation of this one-dimensional linearized system of equations is
\begin{equation}
-i\omega
    \begin{pmatrix}
 \delta \rho_d\\
\delta v_d
\end{pmatrix}
= 
    \begin{pmatrix}
0 & ik\rho_d \\
ik\frac{c_d^2}{\rho_d} & \tau_s^{-1} 
\end{pmatrix}
    \begin{pmatrix}
 \delta \rho_d\\
\delta v_d
\end{pmatrix}
\end{equation}
and the corresponding dispersion relation reads:
\begin{equation}
\label{eq:disp_relation_system3}
    \omega^2-\tau_s^{-1}i\omega+c_d^2k^2=0
\end{equation}
which has the two symmetric solutions
\begin{equation}
\label{eq:solution_disp_relation_system3}
  i\omega_{1,2}=-\frac{1}{2}\tau_s^{-1}\pm\sqrt{\tau_s^{-2}/4-Dk^2\tau_s^{-1}} 
\end{equation}
This solution is plotted in the third column of \autoref{fig:disprelation_nodiff}. On large spatial scales ($k\ll k_c$), the solution of the dispersion relation in equation (\ref{eq:solution_disp_relation_system3}) approaches the drag- and diffusion solutions of the system in which turbulent diffusion is modeled as a source term in the mass equation, i.e., equation (\ref{eq:drag_solution_mass_diff}) and (\ref{eq:diff_solution_mass_diff}). In \autoref{fig:perturbed_solutions}, we plot the numerical solution of this model for $k=0.1 k_c$ at $x = \pi/4k$, corresponding to the drag- and diffusion solution, in the first and second column in red circles. It matches the solution of the model in which diffusion is modeled as a source term in the density equation. However, on spatial scales smaller than the characteristic length scale, the diffusion solution merges with the drag solution and a distinct diffusion solution does not exist anymore in this model. This is clearly visible in \autoref{fig:disprelation_nodiff}. For $k\gg k_c$, the solution to equation (\ref{eq:disp_relation_system3}) approaches 
\begin{equation}
\label{eq:disp_relation_system3_2}
  i\omega_{1,2}=-\frac{1}{2}\tau_s^{-1}\pm i kc_d 
\end{equation} 
In this solution, the effective coupling timescale for drag is increased by a factor of two, i.e., in this model, small-scale perturbations decay on a timescale twice longer compared to the previous model. The physical reason behind the absence of a diffusion solution at small scales ($k\gg k_c$) is the inertia of the diffusion fluxes, which leads to an overshooting of the perturbed states. As a result, perturbations oscillate with frequency $\omega = k c_d$, and propagate with speed $c_d$ as indicated by the imaginary part of equation (\ref{eq:disp_relation_system3_2}). These propagating oscillations, i.e., waves, decay on a timescale equal to twice the stopping timescale rather than the diffusion timescale. On large scales ($k\ll k_c$) these oscillations are overdamped due to drag and, therefore, do not appear in the solution. The wave nature of the solution becomes apparent in the corresponding eigenstates for $k\gg k_c$
\begin{equation}
   \mathrm{Re}(\delta v_{1,2}) = \delta v_0\cos(kx \pm kc_d t)e^{-t/2\tau_s}
\end{equation}
\begin{equation}
    \mathrm{Re}(\delta \rho_{1,2}) =\pm\delta v_0\frac{\rho_0}{c_d\tau_s}\cos(kx \pm kc_d t)e^{-t/2\tau_s}
\end{equation}
which are indeed traveling wave solutions with speed $c_d$ that decay on a timescale twice the stopping time $2\tau_s$. We plot these solutions in the third column of \autoref{fig:perturbed_solutions} at $x=0$ (dotted black lines). The corresponding numerical solution with $k = 10 k_c$ is plotted in red circles. Thus, we numerically confirm the traveling diffusive wave solution theoretically predicted by \cite{Klahr2021}. Although this solution is mathematically valid without restrictions, \cite{Klahr2021} point out the limitations of the model when applied to real physical systems when the diffusion speed $c_d$ approaches the gas sound speed $c_s$. We briefly reiterate their valid arguments here. In reality, the diffusion speed $c_d$ is limited by the actual r.m.s. speed of the dust grains, which is proportional to the diffusivity divided by the correlation time of turbulence \citep[][]{Youdin2007}
\begin{equation}
    v_\mathrm{r.m.s.}^2 = \frac{D}{\tau_t}. 
\end{equation}
Thus, the diffusion speed $c_d$ can be written as a function of the r.m.s. speed
\begin{equation}
    c_d^2 = v_\mathrm{r.m.s.}^2\frac{\tau_t}{\tau_s}
\end{equation}
which, for Kolmogorov turbulence, can be expressed with the Stokes number as 
\begin{equation}
    c_d^2 = v_\mathrm{r.m.s.}^2/St. 
\end{equation}
Thus, mathematically, for strongly coupled grains with $St<1$, the diffusion speed can unphysically exceed the r.m.s. speed of the particles. \newline
Nevertheless, we show that in the diffusive pressure model, the efficiency of turbulent diffusion is limited at small spatial scales due to the inertia of the dust flow, an effect that is not captured in the mass-diffusion model. As a result, there exists a lower limit for the diffusion timescale $t_\mathrm{diff}>2\tau_s$. This makes physical sense because gas turbulence is the underlying driving mechanism of dust diffusions and the dust can not react to gas motions on timescales smaller than the stopping time.

\begin{figure*}
\includegraphics[width=1.95\columnwidth]{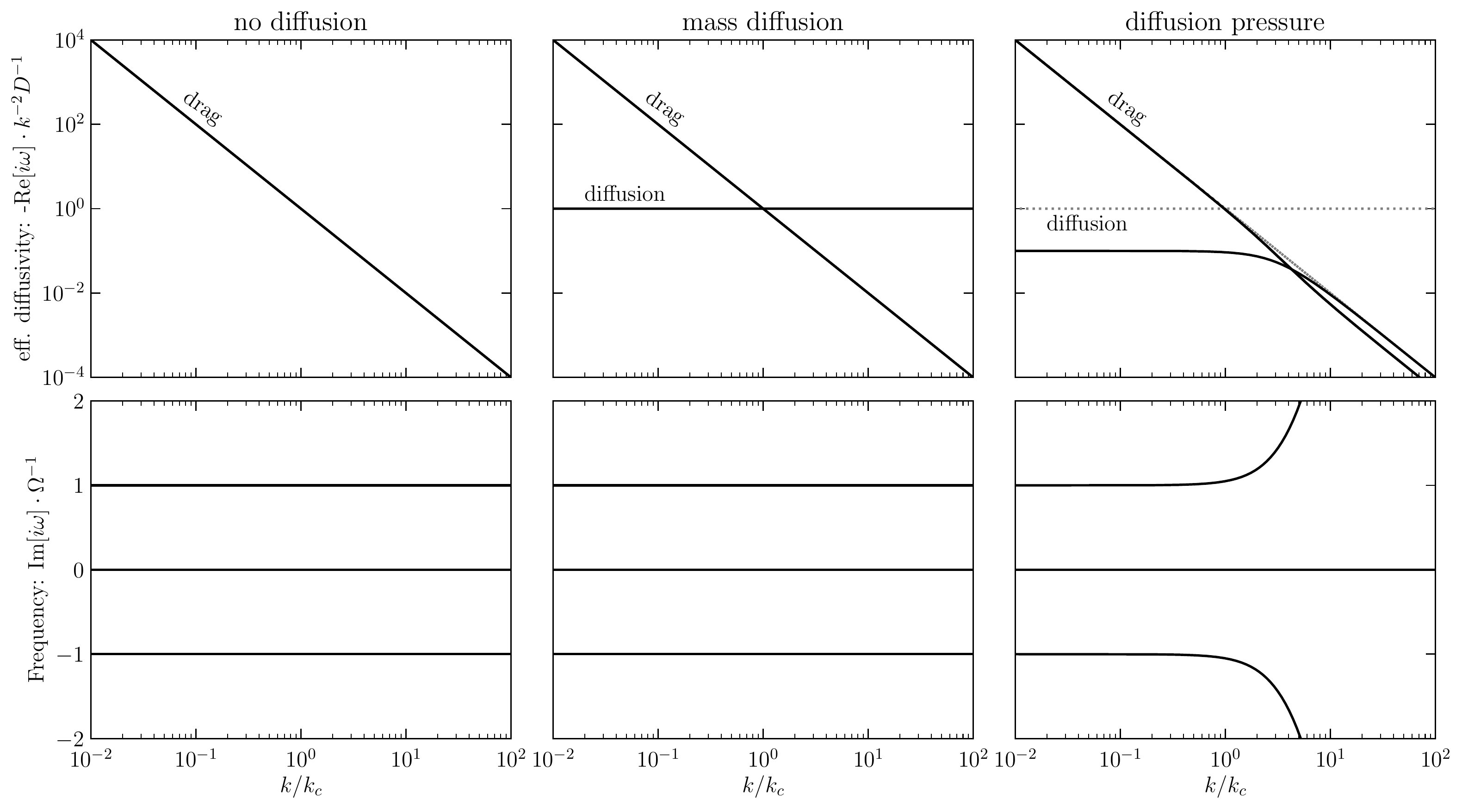}
\caption{Dispersion relations of the two-dimensional linear perturbation analysis in the three turbulent diffusion models, without diffusion (\textit{left}), mass diffusion (\textit{middle}), diffusion pressure (\textit{right}) for $\tau_s = 3 \Omega^{-1}$. The upper row displays the real part of the perturbation frequency $\mathrm{Re}[i\omega]$. The data is multiplied by $k^{-2}$ and normalized by $D^{-1}$. Thus, any diffusive solution is a horizontal line and the drag solution, which is independent of the length scale $k$, has a slope $-2$. The lower row displays the imaginary part of the perturbation frequency $\mathrm{Im}[i\omega]$, normalized by the epicyclic frequency $\Omega$. All three models show oscillating solutions. The two diffusion models differ above the characteristic wave number $k_c$ where the dust-gas coupling sets an upper limit to the diffusivity in the diffusion pressure model. At small wave numbers, the effective diffusivity is decreased by a factor 10 in the diffusion pressure model, in agreement with the $1/(1+St^2)$ behavior. This behavior is not captured in the mass diffusion model.}
\label{fig:disprelation_polar}
\end{figure*}

\begin{figure}
\includegraphics[width=0.9\columnwidth]{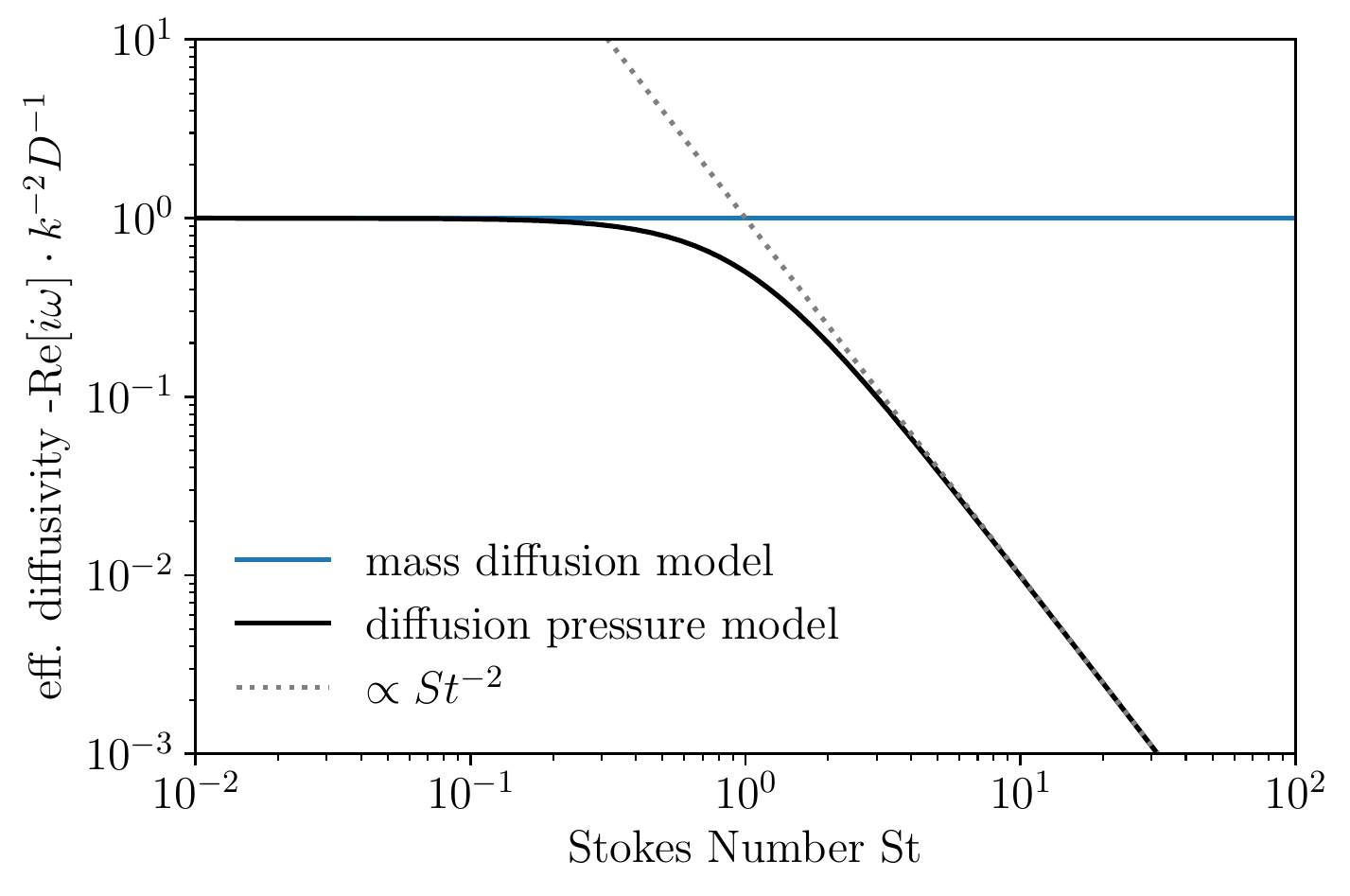}
\caption{Comparison of the effective diffusivity in radial direction vs. the Stokes number between the mass diffusion model (blue) and the momentum conserving diffusion model (black) measured for the diffusion of harmonic perturbations at wave number $k=0.1 k_c$ in an azimuthally symmetric two-dimensional Keplerian disk. As a result of {angular} momentum conservation, the effective diffusivity decreases with $1/(1+St^2)$ 
.}
\label{fig:D_vs_St}
\end{figure}

\begin{figure*}
\includegraphics[width=1.95\columnwidth]{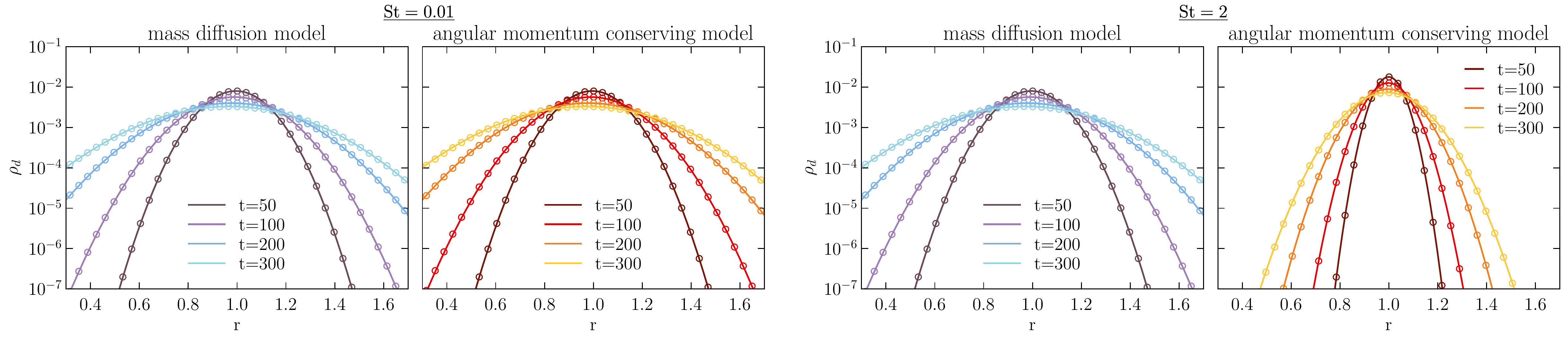}
\caption{Comparison of the radial spreading of an axisymmetric dust ring in cylindrical coordinates between two different diffusion models and at two different Stokes numbers. We set $m=2\cdot 10^{-3}$, $r_0=1$ and $D=10^{-4}$ in code units. In blue color, we show the mass diffusion model. In red color, we show the momentum-conserving model. The solid lines represent the analytic prediction of equation (\ref{eq:diff_spreading}), and the circles represent the numerical solution. The two sub-panels on the l.h.s. show the two models for well-coupled dust ($St=0.01$) for which the two solutions agree. The two sub-panels on the r.h.s. show the two models for moderately coupled dust ($St=2$). The conservation of angular momentum reduces the effective diffusivity by a factor $1/(1+St^2)=5$ compared to the mass diffusion model.}
\label{fig:2D_diffusion_test}
\end{figure*}

\subsection{Two-Dimensional Diffusion in a Keplerian Disk}
\label{sec:Two-Dimensional Diffusion in a Keplerian Disk}
In the one-dimensional analysis, we have ignored external forces and thus also the effects of orbital motions in a Keplerian disk. In this section, we perform the linear perturbation analysis of the systems (\ref{eq:dust_cont_eq2})/(\ref{eq:wo_external_forces}) and (\ref{eq:adv_diff_equation})/(\ref{eq:classical_velocity_equation3}) on an azimuthally symmetric Keplerian disk in two dimensions in polar coordinates. We carry out the perturbation analysis about the radial coordinate $r_0$ and define the dimensionless coordinate $x$ such that $r=r_0(1+x)$ \citep[similarly to][]{Dullemond18}. In a Keplerian disk, the gravitational potential can be expressed as a function of the Keplerian frequency $\Omega$ as $\Phi = -\Omega^2/r^2$. We impose similar harmonic perturbation to the azimuthally symmetric Keplerian disk as in the one-dimensional case. 
\newline
Analogously to section \ref{sec:One Dimension in the Absence of External Forces}, we first analyze the systems (\ref{eq:dust_cont_eq2})/(\ref{eq:wo_external_forces}) and (\ref{eq:adv_diff_equation})/(\ref{eq:classical_velocity_equation3}) in the absence of turbulent diffusion ($D=0$). The resulting linearized equations can be represented by a $3\times 3$ matrix:
\begin{equation}
    -i\omega    
    \begin{pmatrix}
 \delta \rho_d\\
 \delta v_r\\
\delta v_\phi
\end{pmatrix}
=
  \begin{pmatrix}
0 & ik\rho_d & 0 \\
0 & \tau_s^{-1}  & -2\Omega \\
0 & \Omega/2  & \tau_s^{-1} 
\end{pmatrix}
\begin{pmatrix}
 \delta \rho_d\\
 \delta v_r\\
\delta v_\phi
\end{pmatrix}
\end{equation}
which results in the following dispersion relation: 
\begin{equation}
    i\omega\big[(i\omega)^2+2\tau_s^{-1}i\omega+\tau_s^{-2}+\Omega^2\big]=0
\end{equation}
We plot the solution to the dispersion relation in the first column of \autoref{fig:disprelation_polar}. As expected, there exists only a drag solution. As opposed to the one-dimensional solution without external forces, there are two additional solutions which oscillate with the constant epicyclic frequency $\Omega$.\newline
When including turbulent diffusion, the linearized system of equations (\ref{eq:dust_cont_eq2})/(\ref{eq:wo_external_forces}), i.e., the model with pure mass diffusion, with can be written as 
\begin{equation}
    -i\omega    
    \begin{pmatrix}
 \delta \rho_d\\
 \delta v_r\\
\delta v_\phi
\end{pmatrix}
=
  \begin{pmatrix}
Dk^2 & ik\rho_d & 0 \\
0 & \tau_s^{-1}  & -2\Omega \\
0 & \Omega/2  & \tau_s^{-1} 
\end{pmatrix}
\begin{pmatrix}
 \delta \rho_d\\
 \delta v_r\\
\delta v_\phi
\end{pmatrix}
\end{equation}
which has the following dispersion relation: 
\begin{equation}\label{eq:disprel_mass}
\begin{split}
    (i\omega)^3+\big(2\tau_s^{-1}+Dk^2\big)(i\omega)^2+\big(2Dk^2\tau_s^{-1}+\tau_s^{-2}+\Omega^2\big)i\omega\\
    +Dk^2(\tau_s^{-2}+\Omega^2) =0
\end{split}
\end{equation}
The linearized equation of system (\ref{eq:adv_diff_equation})/(\ref{eq:classical_velocity_equation3}), which models turbulent diffusion with the diffusion pressure term, can be written as 
\begin{equation}
    -i\omega    
    \begin{pmatrix}
 \delta \rho_d\\
 \delta v_r\\
\delta v_\phi
\end{pmatrix}
=
  \begin{pmatrix}
0 & ik\rho_d & 0 \\
ikc_d^2/\rho_d & \tau_s^{-1}  & -2\Omega \\
0 & \Omega/2  & \tau_s^{-1} 
\end{pmatrix}
\begin{pmatrix}
 \delta \rho_d\\
 \delta v_r\\
\delta v_\phi
\end{pmatrix}
\end{equation}
where we have replaced the radio of the diffusion coefficient and stopping time with the diffusion speed squared ($c_d^2=D/\tau_s$). The dispersion relation reads:
\begin{equation}\label{eq:disprel_pres}
    (i\omega)^3+2\tau_s^{-1}(i\omega)^2+(\tau_s^{-2}+\Omega_0^2+c_d^2k^2)i\omega+c_d^2k^2\tau_s^{-1} =0
\end{equation}
In \autoref{fig:disprelation_polar}, we plot the solutions to the dispersion relations (\ref{eq:disprel_mass}) and (\ref{eq:disprel_pres}) in the second and third column, respectively for $\tau_s = 3.0\cdot \Omega^{-1}$. Like in the one-dimensional case, we find distinct drag and diffusion solutions also in the two-dimensional case. Moreover, below the characteristic wave number $k_c$, the oscillating epicyclic solutions are also present when turbulent diffusion is included. A new feature present only in the two-dimensional case is the difference in the diffusion solution between the two diffusion models below the characteristic wave number $k<k_c$. There, the effective radial diffusion coefficient in the diffusion pressure model is reduced by a factor of $10$. We analyze this property further by plotting the effective radial diffusion coefficient of the two diffusion models as a function of the Stokes number in \autoref{fig:D_vs_St} at wavenumber $k=0.1 k_c$. The diffusion coefficient in the pure mass diffusion model is independent of the Stokes number. On the other hand, the strength of the effective turbulent diffusion decreases in the diffusion pressure model as the Stokes number approaches unity. In this model, the strength of turbulent diffusion implicitly follows $1/(1+St^2)$ in agreement with \cite{Youdin2007}.

\subsubsection{Radial Turbulent Diffusion in an Axisymmetric Disk}
\label{sec:Radial Turbulent Diffusion in an Axisymmetric Disk}
We aim to numerically confirm the result of the two-dimensional linear perturbation analysis by considering the diffusive spreading of an (initially infinitesimally thin) axisymmetric ring in a two-dimensional Keplerian disk in cylindrical coordinates. \cite{Weber20} derive an analytic solution for this diffusive spreading for the mass diffusion model in their appendix A.3. We summarize the initial assumptions and then list the final solution here. We will expand the initial conditions in the case of the momentum-conserving model to arrive at the same functional expression. \newline
In our test setup, we assume the dust density to be small compared to the dust density ($\rho_d \ll \rho_g$) and the gas density to be constant in time and space with vanishing spatial density gradients ($\nabla\rho_g = 0$). We also assume the diffusion coefficient $D$ to be constant. In the mass diffusion model, the dust is pressureless and, thus, orbits with Keplerian velocity. We also assume the gas to orbit with Keplerian velocity. Thus, the relative velocities between dust and gas vanish ($\mathbf{v}_d-\mathbf{v}_g=0$) and there is no radial drift in the dust due to any potential sub-Keplerian rotation of the gas, only pure diffusive spreading. Thus, in this simplified model, the radial velocities are zero everywhere ($v_{d,r}=v_{g,r}=0$). We further assume the initial dust distribution to be a delta distribution centered around $r_0$ with magnitude $m$ ($\rho_d(r,t=0)=m\delta(r-r_0)$). Then, the density solution has the following form \citep[][]{Weber20}:
\begin{equation}\label{eq:diff_spreading}
    \rho_d(r,t)=\frac{mr_0}{2Dt}\exp \Bigg( -\frac{r^2+r_0^2}{4Dt}\Bigg)I_0\bigg(\frac{r r_0}{2D t}\bigg)
\end{equation}
with $I_0$ being the modified Bessel function of the first kind of order 0. \newline
We also test the diffusive spreading of a ring in the diffusion pressure model. For this, we set up a dust disk that orbits slightly sub-Keplerian to account for the support of the diffusion pressure. We set up the azimuthal gas velocity such that the relative velocity between dust and gas vanish and there is no net radial drift. We initialize the radial component of the dust velocity-based equation (\ref{eq:def_diff_velocity}). We then find the resulting solution to also follow the functional dependency of equation (\ref{eq:diff_spreading}). \newline
In \autoref{fig:2D_diffusion_test}, we compare the diffusive spreading of the two models for well-coupled dust grains (left) and moderately coupled dust grains (right). We set $m=2\cdot 10^{-3}$, $r_0=1$ and $D=1\cdot 10^{-4}$ in code units. The two models produce identical results for well-coupled grains ($St=0.01$). For moderately coupled grains ($St=2$), the radial diffusive spreading proceeds slower in the diffusion pressure model as a result of angular momentum conservation. The effective diffusivity is reduced by a factor $5$ as expected from the predicted $1/(1+St^2)$ behavior. 

%%%%%%%%%%%%%%%%%%%%%%%%%%%%%%%%%%%%%%%%%%%%%%%%%%

% Don't change these lines
\bsp	% typesetting comment
\label{lastpage}
\end{document}